\documentclass[twocolumn,aps,pra,a4paper,superscriptaddress]{revtex4-1}

\usepackage{amsmath}
\usepackage{amssymb}
\usepackage{braket}
\usepackage{graphicx}
\usepackage{verbatim}
\usepackage{amsthm}
\usepackage[usenames]{color}
\usepackage{soul}

\newcommand{\T}[1]{\Theta(#1)}

\date{\today}

\begin{document}
\author{Falko Pientka}
\author{Alessandro Romito}
\author{Mathias Duckheim}
\affiliation{\mbox{Dahlem Center for Complex Quantum Systems and Fachbereich Physik, Freie Universit\"at Berlin, 14195 Berlin, Germany}}
\author{Yuval Oreg}
\affiliation{Department of Condensed Matter Physics, Weizmann Institute of Science, Rehovot, 76100, Israel}
\author{Felix von Oppen}
\affiliation{\mbox{Dahlem Center for Complex Quantum Systems and Fachbereich Physik, Freie Universit\"at Berlin, 14195 Berlin, Germany}}

\title{Signatures of topological phase transitions in mesoscopic superconducting rings}

\begin{abstract}
We investigate Josephson currents in mesoscopic rings with a weak link which are in or near a topological superconducting phase. As a paradigmatic example, we consider the Kitaev model of a spinless $p$-wave superconductor in one dimension, emphasizing how this model emerges from more realistic settings based on semiconductor nanowires. We show that the flux periodicity of the Josephson current provides signatures of the
topological phase transition and the emergence of Majorana fermions
situated on both sides of the weak link even when fermion parity is \emph{not} a
good quantum number. In large rings, the Majorana fermions hybridize only
across the weak link. In this case, the Josephson current is $h/e$
periodic in the flux threading the loop when fermion parity is a good
quantum number but reverts to the more conventional $h/2e$ periodicity in
the presence of fermion-parity changing relaxation processes. In
mesoscopic rings, the Majorana fermions also hybridize through their
overlap in the interior of the superconducting ring. We find that in the
topological superconducting phase, this gives rise to an $h/e$-periodic
contribution even when fermion parity is not conserved and that this
contribution exhibits a peak near the topological phase transition. This
signature of the topological phase transition is robust to the effects of
disorder. As a byproduct, we find that close to the topological phase
transition, disorder drives the system deeper into the topological phase.
This is in stark contrast to the known behavior far from the phase
transition, where disorder tends to suppress the topological phase.
\end{abstract}
\maketitle

\section{Introduction}

On the way to scalable quantum information processing Majorana fermions (MF) in topological superconductors are a promising candidate for the implementation of quantum bits in solid-state devices \cite{Kitaev2001,Nayak2008}. Since information in such systems is stored and processed in a nonlocal fashion by means of their non-Abelian statistics \cite{Read2000,Ivanov2001}, Majorana-based qubits are immune to local fermionic parity conserving perturbations which impair other qubit realizations. Manipulation of such topologically protected qubits requires braiding of MFs, which is well-defined only in two dimensions. However, different schemes have been proposed \cite{Fu2008,Alicea2011,Flensberg2011,Romito2012} to enable braiding of MF in one dimensional systems. Several suggestions for one-dimensional physical realizations that host Majorana bound states (MBS) have been made. These suggestions are based on conventional superconductors in proximity to various systems including a topological insulator edge \cite{Fu2008}, semiconductor wires in a magnetic field \cite{Lutchyn2010,Oreg2010}, and half metals \cite{Duckheim2011,Chung2011}.

Recently, more realistic investigations have elucidat\-ed the effects of interactions and disorder. In general interactions \cite{Gangadharaiah2011,Sela2011,Stoudenmire2011} or disorder with short range cor\-relations \cite{Motrunich2001,Potter2010,Potter2011a,Potter2011b,Brouwer2011,Brouwer2011a,Stanescu2011} can greatly affect the range of parameters in which the system supports topological boundary states or even cause the topological phase to break down completely if the interaction \cite{Gangadharaiah2011} or disorder \cite{Motrunich2001} strength exceed certain critical values. Long-range correlated disorder in a topological superconductor creates nontopological domains with MF localized at the domain walls \cite{Flensberg2010,Shivamoggi2010,Lutchyn2011}.
Several proposals have been put forward to access MFs experimentally based on interferometry \cite{Fu2009a,Hassler2010} or transport properties such as tunneling conductance peak \cite{Law2009,Flensberg2010,Leijnse2011}, half-integer conductance plateaus \cite{Wimmer2011}, or signatures in the shot noise \cite{Bolech2007,Akhmerov2011}.
Recent experiments have reported possible signatures of Majorana bound states in the differential tunneling conductance of semiconductor quantum wires \cite{Mourik2012,Deng2012,Das2012}.

A more specific way of detecting MFs is to measure the Josephson current across a weak link between two topological superconductors \cite{Kitaev2001,Fu2009,Oreg2010,Lutchyn2010,Alicea2011} that arises due to a phase difference of the superconducting order parameters. If the weak link is incorporated into a ring made of a conventional superconductor, the current flowing through the junction is a periodic function of flux with period $h/2e$ (corresponding to $2\pi$ periodicity in the phase difference), associated with the transfer of Cooper pairs across the junction. In a ring made of a topological superconductor, there is a MBS on each side of the junction and the tunneling current obtains a component that is $h/e$ periodic \cite{Akhmerov2011} (corresponding to $4\pi$ periodicity in the phase difference). This doubling of the flux period with respect to the ordinary Josephson effect originates from single-electron tunneling mediated by the MBS and is dubbed fractional Josephson effect.

The $h/e$-periodic Josephson current is observed as long as the fermion number parity of the system is conserved. Once the system is in strict thermodynamic equilibrium, including relaxation processes which change fermion parity, the Josephson current reverts to the conventional $h/2e$ periodicity.  Indeed, the $h/e$-periodic Josephson current has equal magnitude but opposite signs for even and odd fermion parities, so that it averages to zero in the presence of fermion-parity changing processes. Possible workarounds that do not require strict parity conservation rely on the $ac$ Josephson effect \cite{Kwon2004,Jiang2011}, or finite-frequency current noise \cite{Badiane2011}. Experimental signatures of a fractional Josephson effect in Shapiro step measurements have been claimed recently  \cite{Rokhinson2012}.

Here we show that in mesoscopic rings with a weak link, the presence of Majorana fermions can lead to an $h/e$-periodic Josephson current even in thermodynamic equilibrium and in the presence of fermion-parity-breaking relaxation processes. This $h/e$-periodic contribution exists in the topological superconducting phase and peaks in magnitude near the topological phase transition, providing an experimental signature of the phase transition. We investigate this signature for a spinless $p$-wave superconductor wire, the Kitaev chain \cite{Motrunich2001,Kitaev2001}. which is a paradigmatic model exhibiting a topological phase transition. This model also arises as an effective low-energy theory in more realistic situations such as the quantum-wire proposals of Refs.~\onlinecite{Lutchyn2010} and \onlinecite{Oreg2010}. In a ring geometry, the Majorana bound states hybridize not only due to the tunneling across the weak link but also through the superconducting interior of the ring. The latter overlap is exponentially small in the ratio of the ring circumference and the superconducting coherence length governing the spatial extent of the Majorana bound states. As one approaches the topological phase transition, the superconducting coherence length diverges and the interior overlap between the Majorana bound states becomes significant. This causes a peak of the $h/e$-periodic Josephson current near the topological phase transition \footnote{1}.

After discussing this effect in clean rings, we extend our considerations to disordered rings. We show that the signature of the topological phase transition is robust and survives under more realistic conditions. This issue also leads us to study the influence of disorder in the vicinity of the topological phase transition of the Kitaev chain which had not been discussed previously. Previous work \cite{Motrunich2001,Potter2010,Brouwer2011,Brouwer2011a} on disorder effects in the Kitaev chain or models of quantum wires focused on the regime of large chemical potential (measured from the lower band edge), $\mu \gg m\Delta'^2$, where $\Delta'$ denotes the effective $p$-wave order parameter of the Kitaev chain in the continuum limit. In this regime, the topological region in the phase diagram shrinks with increasing disorder \cite{Brouwer2011}. In contrast, the topological phase transition in the Kitaev chain occurs for $\mu=0$ and thus in the opposite regime of $\mu\ll m\Delta'^2$. Remarkably, we find that in this regime disorder {\em increases} the topological region in the phase diagram.

This paper is organized as follows. In Section \ref{sec:model} we review the Kitaev model for a one-dimensional spinless $p$-wave superconductor and its various regimes. We also discuss how this model is related to quantum-wire based realizations, focusing on the modelling of the magnetic flux through a quantum wire ring in proximity to a bulk superconductor. 
Section \ref{sec:signatures} is dedicated to the flux-periodic Josephson currents in clean rings, focusing on the flux-periodicity as a signature of the topological phase transition. The basic effect is discussed in Sec.~\ref{sec:finite_ring}, analytical considerations on the magnitude of the effect are given in Sec.~\ref{sec:low_energy_model}, and a comparison with numerical results is given in Sec.~\ref{sec:numerics}. 
Sec.~\ref{sec:disorder} extends the considerations to disordered rings. Besides a discussion of the effects of disorder on the Josephson currents, we also study the phase diagram of the disordered wire near the topological phase transition.

\section{Model}\label{sec:model}

\subsection{Kitaev model of a one-dimensional spinless $p$-wave superconductor}
\label{sec:kitaev_basics}

Our analysis starts with the Kitaev model of a one-dimensional spinless $p$-wave superconductor \cite{Kitaev2001,Motrunich2001} 
\begin{align}
H_{\rm  TB}&=-\mu_{\rm TB} \sum_{j=1}^N c^\dagger_j c_j\nonumber\\
&-\sum_{j=1}^{N-1} (t c_j^\dagger c_{j+1}+\Delta_{\rm TB} c_jc_{j+1}+\mathrm{h.c.}), 
\label{Kitaev_TB_Hamiltonian}
\end{align}
which describes a wire of $N$ sites. Electrons on site $j$ are annihilated by $c_j$, hop between neighboring sites with hopping amplitude $t$, and have chemical potential $\mu_\mathrm{TB}$. For all numerical calculations in this paper we choose $t=1$. The $p$-wave pairing strength is given by $\Delta_\mathrm{TB}$. Here, we label both the chemical potential and the pairing strength by the subscript TB to distinguish these parameters of the tight-binding model (\ref{Kitaev_TB_Hamiltonian}) from their analogs in the continuum model introduced below. The wire can be closed into a ring with a weak link by an additional hopping term between sites $1$ and $N$, 
\begin{align}
H_{\rm T}=-t' c_N^\dagger c_1 + \mathrm{h.c.}, \label{Kitaev_TB_tunneling_Hamiltonian}
\end{align}
with hopping amplitude $t'$. We assume that charging effects are weak and can be neglected (see Refs.~\onlinecite{Heck2011,Zocher2011} for consequences of charging in ring-like structures).

For an infinite and uniform wire, the Kitaev Hamiltonian (\ref{Kitaev_TB_Hamiltonian}) exhibits a phase transition when the chemical potential $\mu_\mathrm{TB}$ crosses one of the band edges. The system is in a topological (nontopological) superconducting phase when the chemical potential is within (outside) the interval $[-2t,2t]$, i.e., within (outside) the band at vanishing pairing $\Delta_{\mathrm{TB}}=0$. The spectrum exhibits a superconducting gap on both sides of the topological phase transition while the gap closes at the topological critical points $|\mu_\mathrm{TB}| = 2t$. It is thus natural to introduce the chemical potential measured from the lower band edge, i.e., $\mu=\mu_{\mathrm{TB}}+2t$.

In the vicinity of the band edges (say the lower band edge) and thus of the topological phase transition, we can make a continuum approximation to the tight-binding model (\ref{Kitaev_TB_Hamiltonian}).
We will mostly employ the tight-binding model in the first part of the manuscript, while we partially find it more convenient to rely on the continuum approximation in dealing with effects of disorder in Sec.~\ref{sec:disorder}.
The continuum model is formulated in terms of the corresponding Bogoliubov--de Gennes Hamiltonian \cite{Motrunich2001,Kitaev2001}
\begin{align}
H=\begin{bmatrix} \frac{p^2}{2m}+V(x)-\mu & \frac{1}{2}\left\lbrace\Delta'(x) ,p\right\rbrace\\
 \frac{1}{2}\left\lbrace\Delta'(x),p\right\rbrace& -\left(\frac{p^2}{2m}+V(x)-\mu\right) \end{bmatrix} \label{Kitaev_hamiltonian}
\end{align}
where $\Delta'(x)$ is the $p$-wave pairing strength and the curly brackets denote the anticommutator. Here, we have included a disorder potential $V(x)$ which we will return to in more detail in Sec.~\ref{sec:disorder}.  For $V(x)=0 $ the bulk spectrum of the continuum model is given by 
\begin{align}
\epsilon_p=\pm\left[ \left(\frac{p^2}{2m}-\mu\right)^2+|\Delta'|^2 p^2\right]^{1/2} , \label{Kitaev_spectrum}
\end{align}
which becomes gapless for $\mu=0$. This indicates the above-mentioned topological phase transition between a topological phase with $\mu>0$ and 
a nontopological phase for $\mu<0$.
\begin{figure}[t!]
 \begin{center}
\includegraphics[width=.23\textwidth]{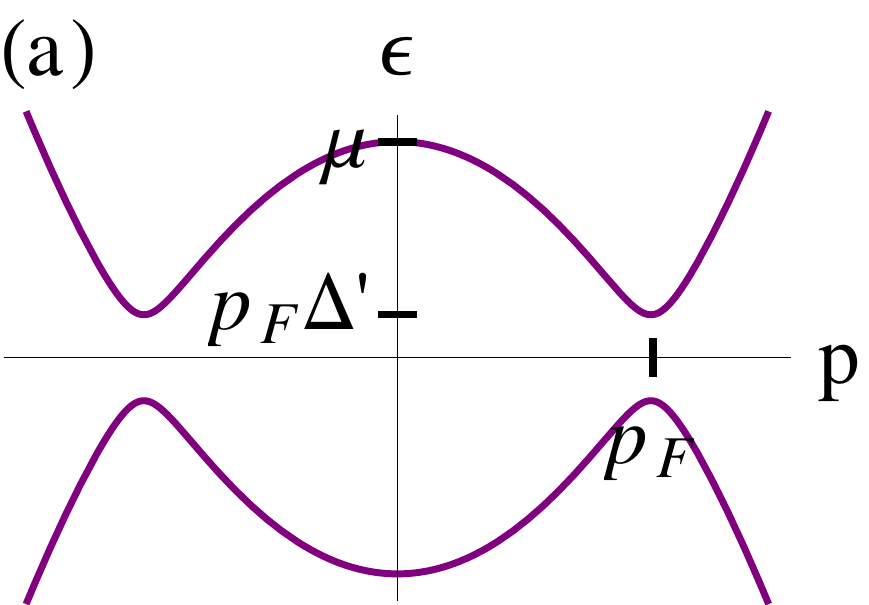}
\includegraphics[width=.23\textwidth]{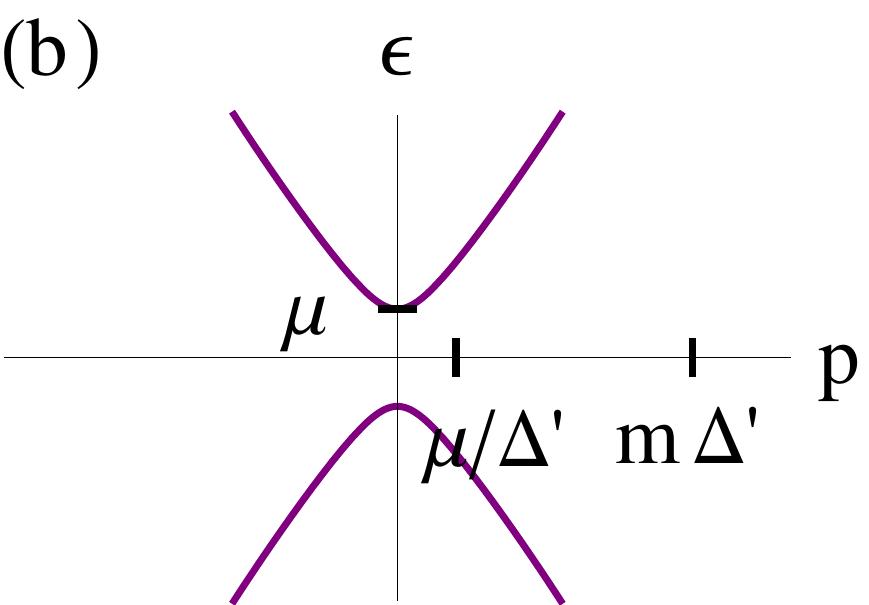}
\end{center}
\caption{(Color online) Bulk spectrum Eq.~(\ref{Kitaev_spectrum}) of Kitaev's model for a spinless $p$-wave superconductor in the regimes (a) $\mu\gg m\Delta'^2$ and (b) $0<\mu\ll m\Delta'^2$.}
\label{fig:spectrum}
\end{figure}

In a semi-infinite wire, the topological phase is characterized by a Majorana bound state localized near its end point. The Majorana bound state has zero energy and a wave function that decays exponentially into the wire on the scale of the superconducting coherence length $\xi$. In a finite wire, the Majorana bound states localized at the two ends of the wire hybridize and form a conventional Dirac fermion whose energy $\epsilon_0$ scales like the overlap of the two Majorana end states which is exponentially small in the length $L$ of the wire. The wavefunction of the Majorana bound state depends on the parameter regime (see, e.g., Ref.~\onlinecite{Halperin2012}). This is easily seen by determining the allowed wavevectors at zero energy from Eq.~(\ref{Kitaev_spectrum}), which yields
\begin{align}
p_0& =\pm im|\Delta'|\pm \sqrt{2m\mu-m^2|\Delta'|^2}.\label{MBS_wavevector}
 \end{align}

(i) $\mu\gg m\Delta'^2$: Deep in the topological phase, the bulk excitation spectrum Eq.~(\ref{Kitaev_spectrum}) has two minima around $\pm p_F=\pm\sqrt{2m\mu}$ with a gap $\Delta_{\rm eff}^{(i)}\approx p_F\Delta'$ (see Fig.~\ref{fig:spectrum}a). According to Eq.~(\ref{MBS_wavevector}), the Majorana wavefunctions decay on the scale $\xi=1/m\Delta'$ and oscillate with a much shorter period $1/p_F$. 
In a finite wire the hybridization energy is given by (cf.\ appendix~\ref{appendice}) $\epsilon_0= 2\Delta'p_F|\sin(p_F L)|\exp(-L/\xi)$, which has accidental degeneracies at integer values of $p_FL/ \pi$.

(ii) $\mu\ll m\Delta'^2$: Near the topological phase transition at $\mu=0$, the excitation spectrum has only a single minimum at $p=0$ with a gap of order $\mu$ (see Fig.~\ref{fig:spectrum}b). At low energies, we can neglect the kinetic energy in Eq.~(\ref{Kitaev_hamiltonian}) and the spinless $p$-wave superconductor can be approximately described by the Dirac Hamiltonian 
\begin{align}
H\simeq -\mu\tau_z+\Delta' p\tau_x. \label{Dirac_Hamiltonian}
\end{align}
Eq.~(\ref{MBS_wavevector}) gives $p_0\approx \pm i\mu/\Delta'$, so that the spatial extent of the Majorana wavefunction is governed by the coherence length $\xi=\Delta'/\mu$, which diverges at the topological phase transition. In contrast to the previous regime, the end-state energy does not exhibit oscillations, $\epsilon_0 \propto \exp(-L/\xi)$.

\subsection{Magnetic flux}\label{sec:magnetic_flux}
\begin{figure}[t!]
 \begin{center}
\includegraphics[width=0.48\textwidth]{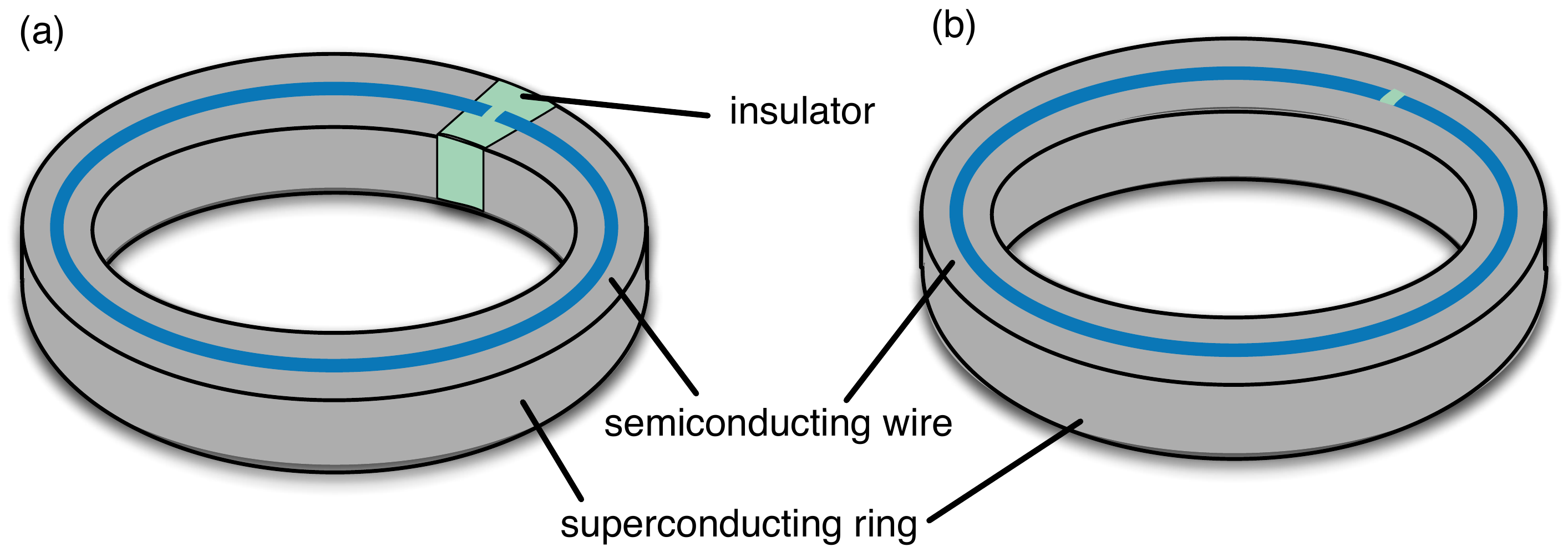}
\end{center}
\caption{(Color online) Two possible setups for a quantum wire with a tunneling junction in proximity to an $s$-wave superconductor. (a) The bulk superconductor is interrupted by an insulating region underneath the weak link in the wire. (b) The bulk superconductor forms a continuous ring and only the wire contains a weak link.}
\label{fig:setup}
\end{figure}

In the presence of a magnetic flux threading the ring, both the tunneling amplitude and the pairing strength become complex and acquire a phase. The precise nature of these phases depends on the physical realization of the Kitaev chain. We illustrate this point by discussing two possible setups based on the proposal to realize the Kitaev chain in a semiconductor wire proximity coupled to an $s$-wave superconductor \cite{Lutchyn2010,Oreg2010}, as illustrated in Fig.~\ref{fig:setup}:
\begin{enumerate} 
\item[(a)] The $s$-wave superconductor is interrupted underneath the weak link in the quantum wire. Current can flow around the loop only through the semiconductor weak link. 
\item[(b)] The $s$-wave superconductor forms a closed ring and a weak link exists only in the semiconductor wire. The current through the weak link of the semiconductor will in general be only a small perturbation of the current flowing through the superconductor.  
\end{enumerate}

We assume that the thickness of the superconducting ring is small compared to both its London penetration depth and its superconducting coherence length $\xi_{\mathrm{SC}}$. The supercurrent flowing in the superconductor is given by \cite{Tinkham1975}
\begin{align}
 J_s=\frac{2e}{m^*}|\psi|^2\left(\hbar\nabla\varphi-2 eA\right),\label{supercurrent}
\end{align}
where $m^*$ and $|\psi|^2$ are the effective mass and density of the superconducting electrons and $\varphi$ denotes the phase of the $s$-wave order parameter. The $p$-wave pairing potential in the quantum wire inherits its phase $\varphi$ from the $s$-wave superconductor underneath via the proximity effect. (The effective $p$-wave order parameter may have an additional phase shift that depends on geometric details such as the direction of the Zeeman field and the spin-orbit coupling; however, these contributions lead to constant offsets of the phase which are unaffected by the magnetic flux.) The vector potential ${\bf A}$ oriented along the wire is related to the Aharonov--Bohm flux $\phi$ through
\begin{align}
\phi=\oint\mathrm{d}x A(x)\label{stokes_theorem},
\end{align}
where the integral is taken around the ring of circumference $L$.

The phase of the order parameter $\varphi$ is different for the two setups illustrated in Fig.~\ref{fig:setup}.
In setup (a), no supercurrent is able to flow since the loop is interrupted, $J_s=0$. If we choose a gauge in which the vector potential is uniform around the ring, $A(x) = \phi/L$, the phase $\varphi$ of the order parameter becomes $\varphi (x) =  4\pi(\phi/\phi_0)(x/L)$ in terms of the normal-metal flux quantum $\phi_0 = h/e$. 
In setup (b), the supercurrent around the ring is governed by fluxoid quantization $\varphi (x+L) = \varphi(x) + 2\pi n$, with the integer $n$ labeling the fluxoid states. In a gauge in which $A(x)=\phi/L$, this implies that $\nabla \varphi = 2\pi n/L$, yielding a supercurrent of $J_s = (2e/m^*)|\psi|^2 [2\pi \hbar n/L - 2e A]$. Here, $[x]$ denotes the integer closest to $x$.  In thermodynamic equilibrium, the system realizes the fluxoid state of lowest energy and thus of lowest supercurrent, i.e., $n=[\phi/(\phi_0/2)]$. 

Within the chosen gauge, in setup (a) the hopping amplitude and the pair potential in the tight binding Hamiltonian in Eq.\ (\ref{Kitaev_TB_Hamiltonian}) take the form $t \to t e^{i 2\pi \phi/N\phi_0}$ and $\Delta_\mathrm{TB} \to \Delta_\mathrm{TB}e^{i 4\pi (\phi/\phi_0) (j/N)}$. 
Alternatively one can perform the gauge transformation  $c_j \to c_j e^{-i(j-1/2) 2 \pi \phi/N \phi_0}$ which eliminates the phase from the pair potential. In this new gauge, both the pair potential and the hopping amplitude $t$ in the interior of the ring are real while the hopping amplitude across the weak link acquires a phase factor,
$t^\prime \to t^\prime e^{i2\pi \phi/\phi_0}$. Our numerical results will be obtained for this representation of the tight-binding model. 
In contrast, in setup (b), we find $\Delta_\mathrm{TB} \to \Delta_\mathrm{TB}e^{i 2\pi [\phi/(\phi_0/2)] (j/N) }$ for the pair potential (notice the closest integer symbol $[.]$ in the exponent), while $t \to t e^{i 2\pi \phi/N\phi_0}$ as well as $t^\prime \to t^\prime e^{i 2\pi \phi/N\phi_0}$. As in the previous case (a), we can eliminate the phase of the pair potential by a gauge transformation. However, this no longer eliminates the phase of the hopping matrix element $t$. Instead, one finds $t \to t e^{i (\pi/N) \{ \phi/(\phi_0/2) - [\phi/(\phi_0/2)] \} }$ and $t^\prime \to t^\prime e^{i (\pi/N) \{ \phi/(\phi_0/2) +(N-1) [\phi/(\phi_0/2)]\} }$. The fact that we can no longer eliminate the magnetic flux from the bulk of the wire is a manifestation of the fact that supercurrents in the $s$-wave superconductor modify the spectrum of the quantum wire \cite{Romito2012}.
 
Clearly, the effective Kitaev chain is quite different for setups (a) and (b). In the remainder of this paper, we will focus on setup (a) where the flux enters only into the tunneling Hamiltonian representing the weak link. In this setting, the current in the semiconductor wire of interest here is experimentally more accessible since there is no  background current in the bulk $s$-wave superconductor unlike in setup (b).

\section{Clean rings}\label{sec:signatures}
\subsection{Infinite wire}

We first briefly review the Josephson effect of two semi-infinite wires connected at their ends through a weak link (or equivalently, a ring of infinite circumference), as originally considered by Kitaev \cite{Kitaev2001}.
 The corresponding low-energy excitation spectrum as a function of flux is sketched in Fig.~\ref{fig:flux}a. Due to the Majorana end states, there are two subgap states whose energies are governed by the tunneling amplitude across the weak link. While each individual level is periodic in flux with period $h/e$, the overall spectrum is $h/2e$ periodic. As a result, the thermodynamic ground state energy of the system -- and thus the Josephson current in strict thermodynamic equilibrium -- are $h/2e$ periodic.

At the same time, the $h/e$ periodicity of the individual subgap states is a direct consequence of the Majorana nature of the endstates. This signature of Majorana fermions can be brought out in measurements of the Josephson current if the fermion parity of the system is a good quantum number. The level crossing of the two Majorana subgap states in Fig.~\ref{fig:flux}a is then protected by fermion parity conservation. As a result, since there is only a single level crossing per superconducting flux quantum, the system necessarily goes from the ground state to an excited state (or vice versa) when changing the flux by $h/2e$. During this process, the excited state is unable to relax to the ground state since this would require a change in fermion parity. Thus, the system only returns to its initial state after a change in flux of $h/e$, which corresponds to the fractional Josephson effect.

\subsection{Finite size ring}\label{sec:finite_ring}

\begin{figure}[t!]
\includegraphics[width=.24\textwidth]{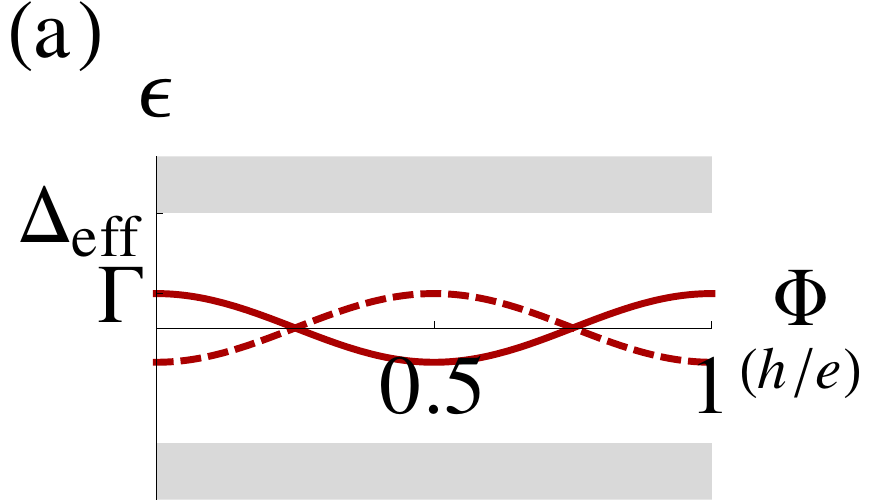}
\includegraphics[width=.23\textwidth]{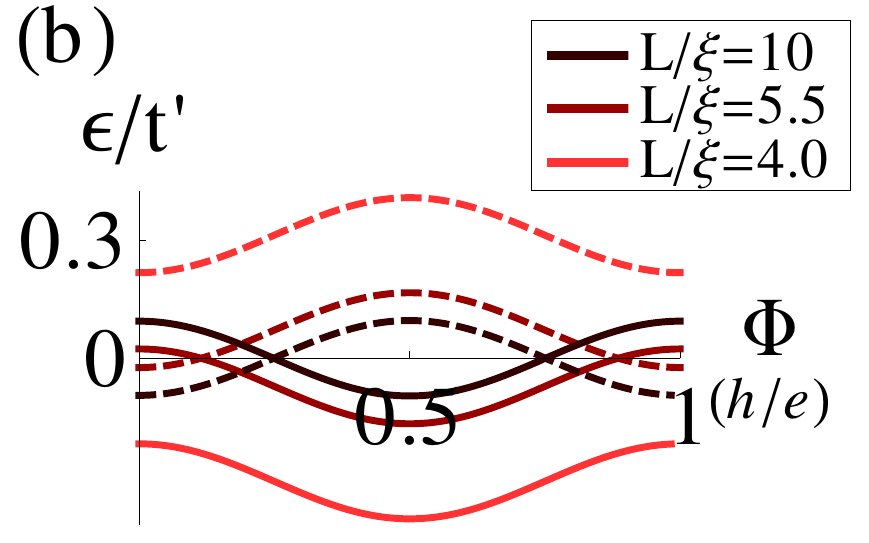}
\caption{(Color online) (a) Typical Bogoliubov--de Gennes spectrum as a function of phase difference across the junction between two semi-infinite wires in the topological phase with the tunneling amplitude $\Gamma$ and the gap $\Delta_{\rm eff}$. The two low-energy Majorana states represented by the dashed and solid lines are related by particle-hole symmetry. The continuum of states outside the gap is displayed in gray. The thermodynamic ground state has period $h/2e$.
(b) Numerical results for the subgap spectrum of a mesoscopic ring with finite circumference for $\Delta=1$, $\mu=-1.8$, $t'=0.01$. We set $t=1$ for all numerical calculations in this paper. The parameters correspond to $\xi=9.5$ and the different curves display data for ring circumferences $L=95,52,38$ all in units of the lattice spacing.
As the circumference of the wire decreases the overlap through the topological superconductor in Eq.~(\ref{low_energy_Hamiltonian}) increases. Note that the equilibrium ground state always has $h/e$ periodicity in rings of finite circumference.}
\label{fig:flux}
\end{figure}

For rings with finite circumference, the two Majorana bound states localized at the two banks of the weak link hybridize not only through the tunnel coupling across the weak link but also because of the overlap of their wavefunctions in the topological superconductor. In the previous subsection, we considered the situation in which this interior hybridization is vanishingly small compared to tunneling across the weak link. Conversely, when tunneling across the weak link is negligible compared to the interior hybridization, the splitting of the Majoranas due to the interior overlap does not depend on flux. Weak tunneling across the junction will then cause a small $h/e$-periodic modulation of the split Majorana levels with flux. In this situation, even the {\em thermodynamic} ground state energy becomes $h/e$ periodic, regardless of the presence or absence of fermion parity violating processes. In fact, of the two $h/e$-periodic levels, the negative-energy level (which is occupied in equilibrium) corresponds to an even-parity ground state while the positive-energy level is occupied in the odd-parity first excited state. Weak fermion parity violating processes will not destroy the $h/e$-periodic Josephson current as the two levels no longer cross as function of flux. 

The full crossover of the Bogoliubov--de Gennes spectrum as the interior overlap of the Majorana bound states increases is illustrated with numerical results in Fig.~\ref{fig:flux}b  (see Sec.~\ref{sec:numerics} for details on the numerical calculations). They confirm the above picture for the limit of strong overlap. But they also show that an $h/e$-periodic contribution to the equilibrium Josephson current exists even when the interior splitting is of the order of or smaller than the tunnel coupling across the weak link. Indeed, the interior overlap essentially pushes one of the two states (dashed line) up in energy, while it pushes its particle-hole conjugate state (solid line) down. At small interior overlaps, this shifts the two level crossings (initially at $\phi_0/4$ and $3\phi_0/4$) outwards towards a flux of zero and one flux quantum $\phi_0$. Note that the level crossings remain intact, protected by fermion parity conservation. However, once the level crossings reach a flux of zero and $\phi_0$, respectively, the levels merely touch at these points. Thus, fermion parity no longer protects the levels from splitting, and indeed one state remains at finite and negative energies at all values of flux while, symmetrically, its particle-hole conjugate state remains at finite and positive energies. 

Consider now the Josephson current as function of flux in the presence of weak but finite fermion parity violating processes. Specifically, we assume that the flux is varied by $h/e$ on a time scale 
which is large compared to the relaxation time of the fermion parity while at the same time, the fermion parity violating processes are weak compared to the hybridization of the Majorana bound states so that the Bogoliubov-de Gennes spectra in Fig.~\ref{fig:flux} are relevant. In this case, the Josephson current is essentially $h/2e$ periodic deep in the topological phase, where $L\gg \xi$. However, as the system approaches the topological phase transition, $\xi$ grows and hence, the hybridization of the Majorana bound states through the interior of the ring increases. As a result, the $h/e$-periodic contribution to the current increases. Conversely, the Majorana bound states disappear on the nontopological side of the phase transition where the Josephson current thus reverts to $h/2e$ periodicity. As a result, we expect a {\em peak} in the $h/e$-periodic Josephson current near the topological phase transition, whose measurement would constitute a clear signature of the topological phase transition and the formation of Majorana fermions.

\begin{figure}[tp!]
\includegraphics[width=.48\textwidth]{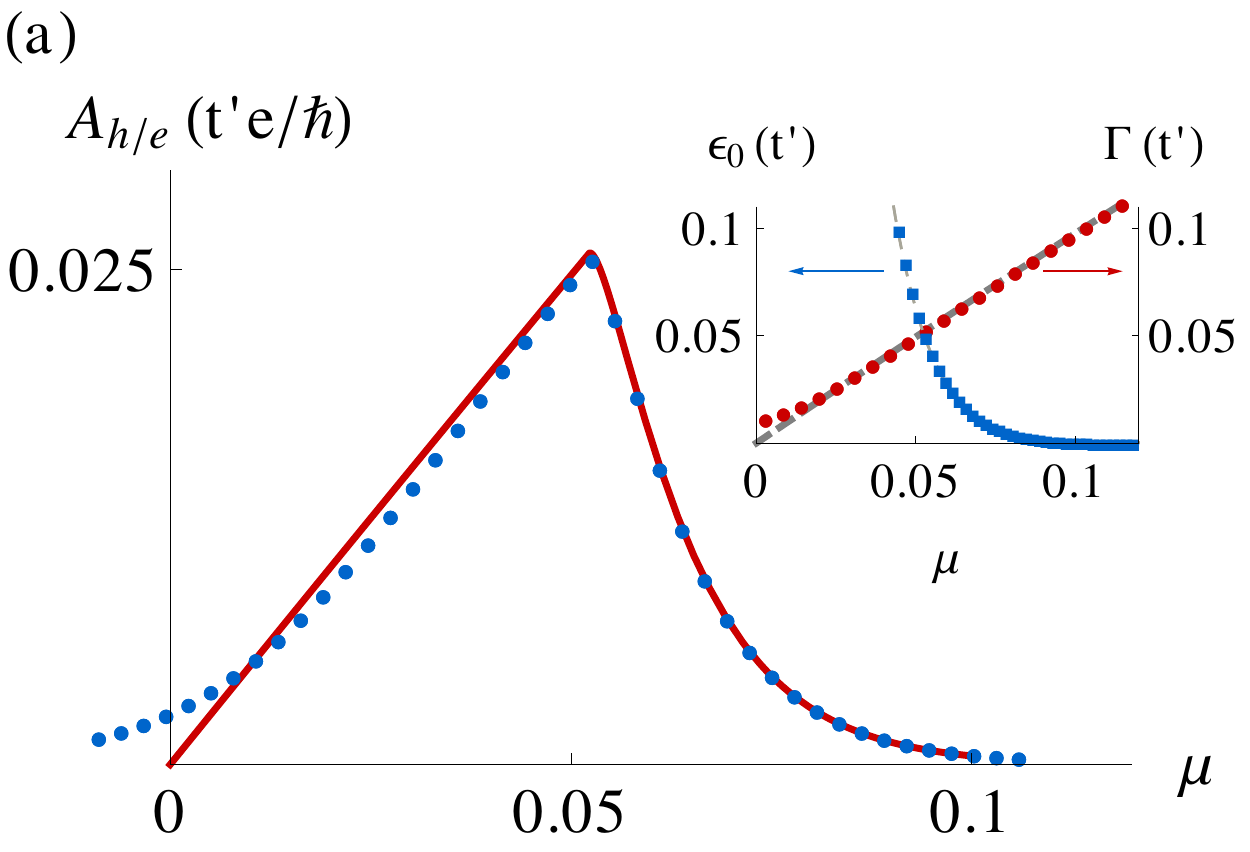}\\
\includegraphics[width=.48\textwidth]{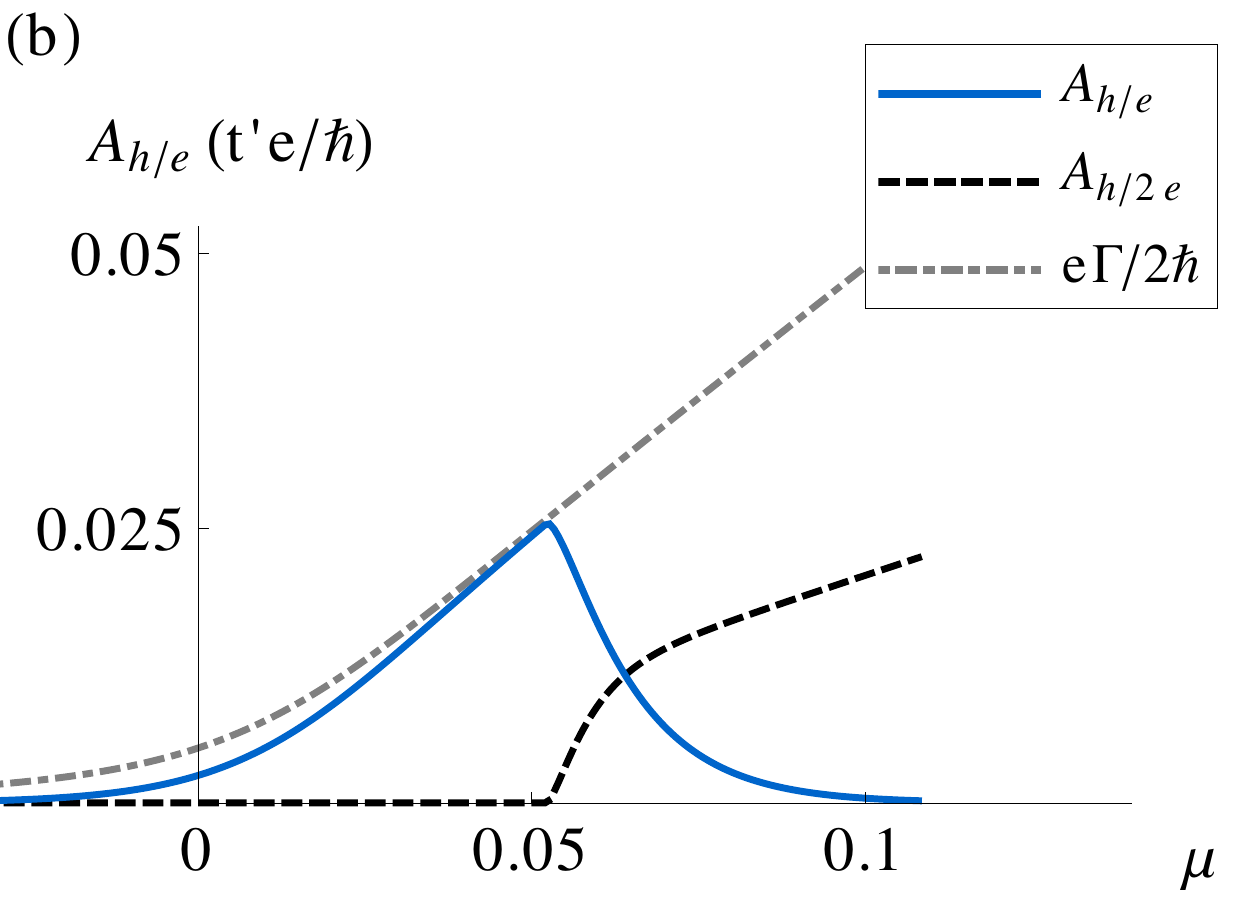}\\
\includegraphics[width=.48\textwidth]{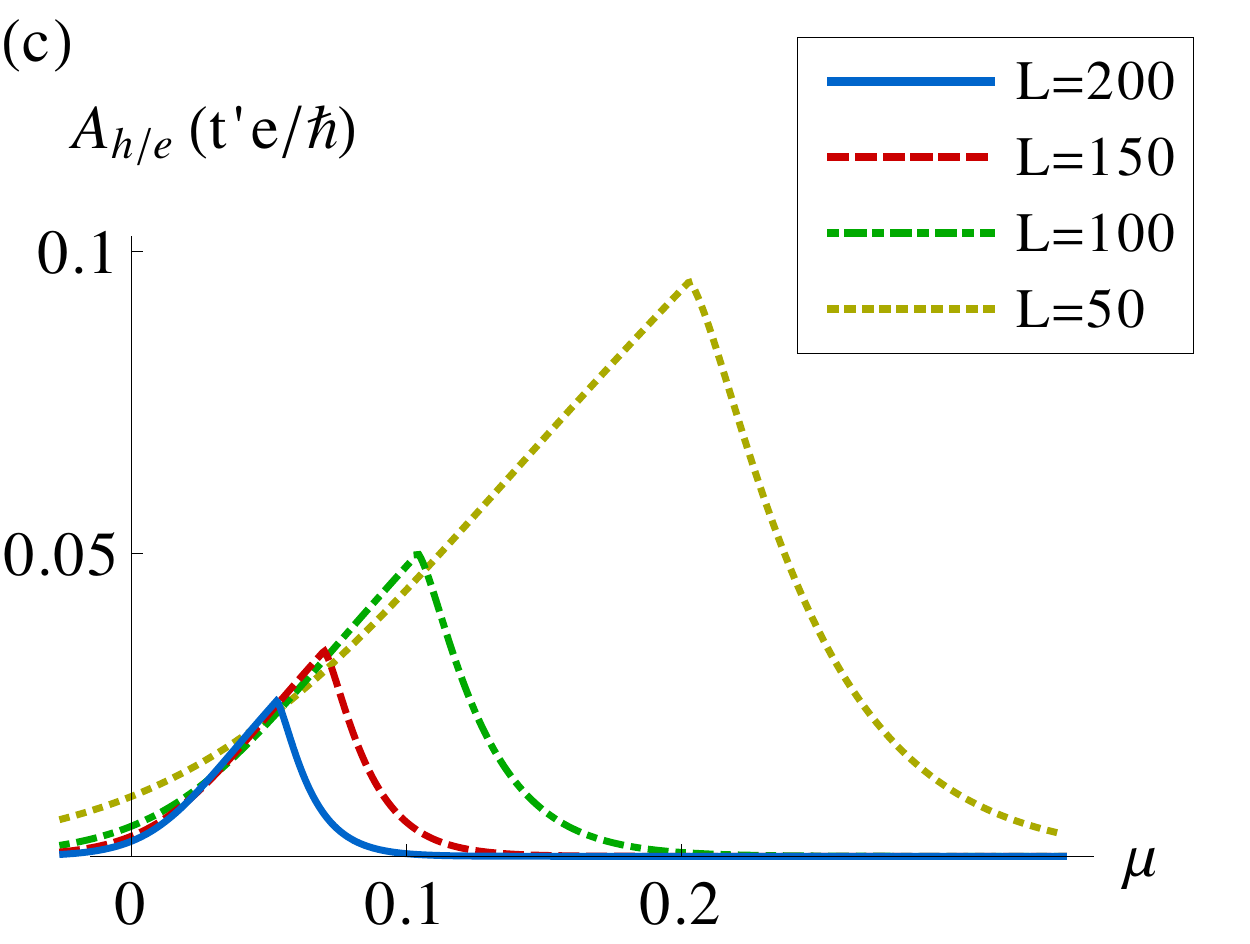}
\caption{(Color online) (a) Numerical results for the $h/e$-periodic Fourier component of the Josephson current, $A_{h/e}$, as a function of chemical potential  in a ring with $\Delta_{\rm TB}=1$ and $L=200$ (blue dots) together with analytical expression Eqs.~(\ref{fourier}) (red solid line).
Inset: numerical results for $\epsilon_0$ (blue squares) and $\Gamma$ (red circles) together with the corresponding analytical expressions (gray dashed curves) Eqs.~(\ref{energy-splitting}) and (\ref{gamma_mean}).
(b) $h/e$-periodic Fourier component (solid), $h/2e$-periodic Fourier component (dashed), and the maximum tunneling current of the MBS, $e\Gamma/h$. 
(c) $A_{h/e}$ for different ring circumferences $L$. }
\label{fig:fourier_component}
\end{figure}

This expectation is confirmed by the numerical results shown in Fig.~\ref{fig:fourier_component}a, where the corresponding Fourier coefficient $A_{h/e}=(2e/h)\int_0^{h/e}\mathrm{d}\phi I(\phi)\sin(2\pi e\phi/h)$ of the {\em equilibrium} Josephson current $I (\phi)$ is plotted as a function of chemical potential. $A_{h/e}(\mu)$ exhibits a peak in the topological phase ($\mu>0$), which moves closer to the topological phase transition at $\mu=0$ as the ring circumference increases (see Fig.~\ref{fig:fourier_component}b). 

Deep in the topological phase the Majorana bound states are localized at the weak link. Approaching the phase transition at $\mu=0$, the MBS delocalize. On the one hand, this causes an increase in the overlap of the MBS in the interior of the topological superconductor. As discussed above, this leads to an increase of the $h/e$-periodic Josephson current. On the other hand, however, the probability density of the Majorana bound state near the weak link decreases, causing a suppression of the hybridization of the Majorana bound states across the weak link and hence of the $h/e$-periodic Josephson current. Thus, the peak occurs for the value of $\mu$ where the interior overlap splitting is equal to the tunnel coupling. Since the interior overlap is exponentially small in $L/\xi$ while the hybridization across the weak link is roughly independent of the ring's circumference $L$, the peak position shifts towards the phase transition point at $\mu=0$ with increasing $L$ (cf.\ Fig.~\ref{fig:fourier_component}c). Since at the same time the $h/e$-periodic Josephson current becomes suppressed when the systems is approaching the phase transition, the peak is more pronounced in shorter rings.

\subsection{Low-energy Hamiltonian} \label{sec:low_energy_model}

A more quantitative description can be developed by restricting the Hamiltonian to the low-energy subspace spanned by the two Majorana bound states. The pro\-jec\-tion of the tunneling Hamiltonian across the weak link onto this subspace yields 
\begin{align}
H_{\rm T}=-\Gamma\cos(2\pi \phi/\phi_0) (d_{\rm M}^\dagger d_{\rm M}-1/2), \label{effective_tunneling_hamiltonian}
\end{align}
where $d_M$ is the Dirac fermion constructed from the two Majorana bound states. The parameter $\Gamma$ measures the tunnel coupling of the Majorana bound states across the weak link and is given by (cf.\ Eq.~(\ref{good-for-plots}) in appendix~\ref{appendice})
\begin{align}
 \Gamma=\frac{ t'\mu (4t-\mu) \Delta_{\mathrm{TB}} }{ t(t+\Delta_{\mathrm{TB}})^2} \label{gamma_mean}.
\end{align}
Here the factor of $\mu$ accounts for the probability density of the Majorana wavefunction at the junction, which vanishes at the phase transition.

The overlap of the Majorana end-states in the interior of the wire leads to an additional coupling (cf.\ appendix~\ref{appendice}) 
\begin{align*}
H_{\mathrm{overlap}}=\epsilon_0 \left(d_M^{\dag} d_M-1/2\right),
\end{align*}
where
\begin{equation}
\epsilon_0 =2 \mu \exp(-L/\xi)
\label{energy-splitting}
\end{equation}
measures the strength of the overlap.

Combining these two contributions for a mesoscopic ring near the topological phase transition ($\mu\ll m{\Delta^\prime}^2$), the effective low-energy Hamiltonian reads as
\begin{align}
 H_{\rm eff}&=\left[\epsilon_0-\Gamma\cos\left(\frac{2\pi\phi}{\phi_0}\right)\right]\left(d_{\rm M}^\dagger d_{\rm M}-1/2\right).\label{low_energy_Hamiltonian}
\end{align}
The Bogoliubov--de Gennes spectrum of this Hamiltonian reproduces the numerically calculated subgap spectra depicted in Fig.~\ref{fig:flux}b.

In principle, both the negative energy continuum states as well as the negative energy subgap state contribute to the equilibrium Josephson current. If we denote the sum over all negative excitation energies by $E_0(\phi)$, we can write the equilibrium Josephson current as $I(\phi)=-\partial_\phi E_0(\phi)$.
However, it is natural to expect and will be coroborated by our numerical results that the Josephson current is dominated by the contribution of the subgap state $I(\phi)\simeq \partial_\phi|\epsilon_0-\Gamma\cos(2\pi\phi/\phi_0)|/2$. Thus, it is straight-forward to compute the $h/e$-periodic Fourier component of the Josephson current,
\begin{align}
A_{h/e} = \left\lbrace
\begin{matrix}\frac{e\Gamma}{\pi \hbar}  \left[\frac{\epsilon_0}{\Gamma}\sqrt{1-\frac{\epsilon_0^2}{\Gamma^2}} + \arcsin\left( \frac{\epsilon_0}{\Gamma} \right) \right], & \epsilon_0 <\Gamma \\
\frac{e\Gamma}{2\hbar},& \epsilon_0 >\Gamma\end{matrix} \right. .
\label{fourier}
\end{align}
In the next section, we compare this analytical result with numerics and find nice agreement.

\subsection{Numerical Results}\label{sec:numerics}

To obtain numerical results for the Josephson current, we solve the Hamiltonian defined in Eqs.~(\ref{Kitaev_TB_Hamiltonian}) and (\ref{Kitaev_TB_tunneling_Hamiltonian}) by exact diagonalization. Fig.~\ref{fig:fourier_component}a compares the amplitude of the $h/e$-periodic component as a function of chemical potential with the analytical result in Eq.~(\ref{fourier}). The numerical results agree well with the behavior predicted by the low-energy model, except for small deviations in the immediate vicinity of the phase transition at $\mu=0$. In the inset of Fig.~\ref{fig:fourier_component}a we compare the analytical and numerical results for the quantities $\Gamma$ and $\epsilon_0$ appearing in the low-energy Hamiltonian. While the model correctly captures $\epsilon_0$ in the regime of interest, there are deviations of $\Gamma$ near $\mu=0$. These discrepancies are readily understood as a consequence of the finite circumference of the ring. Although the coherence length diverges at the phase transition, 
the Majorana bound states can delocalize at most throughout the entire length of the ring 
there remains a finite probability density of the Majorana bound state wavefunction at the weak link. 

Figure~\ref{fig:fourier_component}b shows that the left flank of the peak of $A_{h/e}$ and $e\Gamma/h$ deviate slightly. This deviation is a measure of the size of the bulk contribution to the $h/e$-periodic current. The latter can thus be seen to be small, justifying our focus on the low-energy Hamiltonian (12) describing the Majorana bound states only. In the same figure, the $h/2e$ component is plotted, showing that the $h/e$-periodic Josephson current exceeds the $h/2e$ component. This is a consequence of the tunneling regime that favors single-electron tunneling over the tunneling of Cooper pairs.

In Fig.~\ref{fig:fourier_component}c, we show how the position of the peak in $A_{h/e}$ depends on the circumference of the ring. We find that the value of $\mu$ where the peak occurs scales as $1/L$. This result can be understood as follows. $\Gamma$ is essentially independent of the length of the ring, while $\epsilon_0$ scales as $\sim\exp(-L/\xi)$. As we have seen above the peak occurs at  $\epsilon_0=\Gamma$. For given $t$, $\Delta_{\rm TB}$, and $t'$, $\Gamma$ is fixed and the peak occurs at a constant value of the ratio $L/\xi$. Since $\xi \sim 1/\mu$, the value of $\mu$ where the peak occurs scales as $1/L$.
Also note that the above-mentioned tail of the peak at $\mu\leq 0$ originating from finite-size corrections is more pronounced in shorter rings.
 
\section{Effects of disorder}\label{sec:disorder}

\subsection{$h/e$-periodic Josephson current in disordered rings}

\begin{figure}[tp!]
 \begin{center}
\includegraphics[width=.46\textwidth]{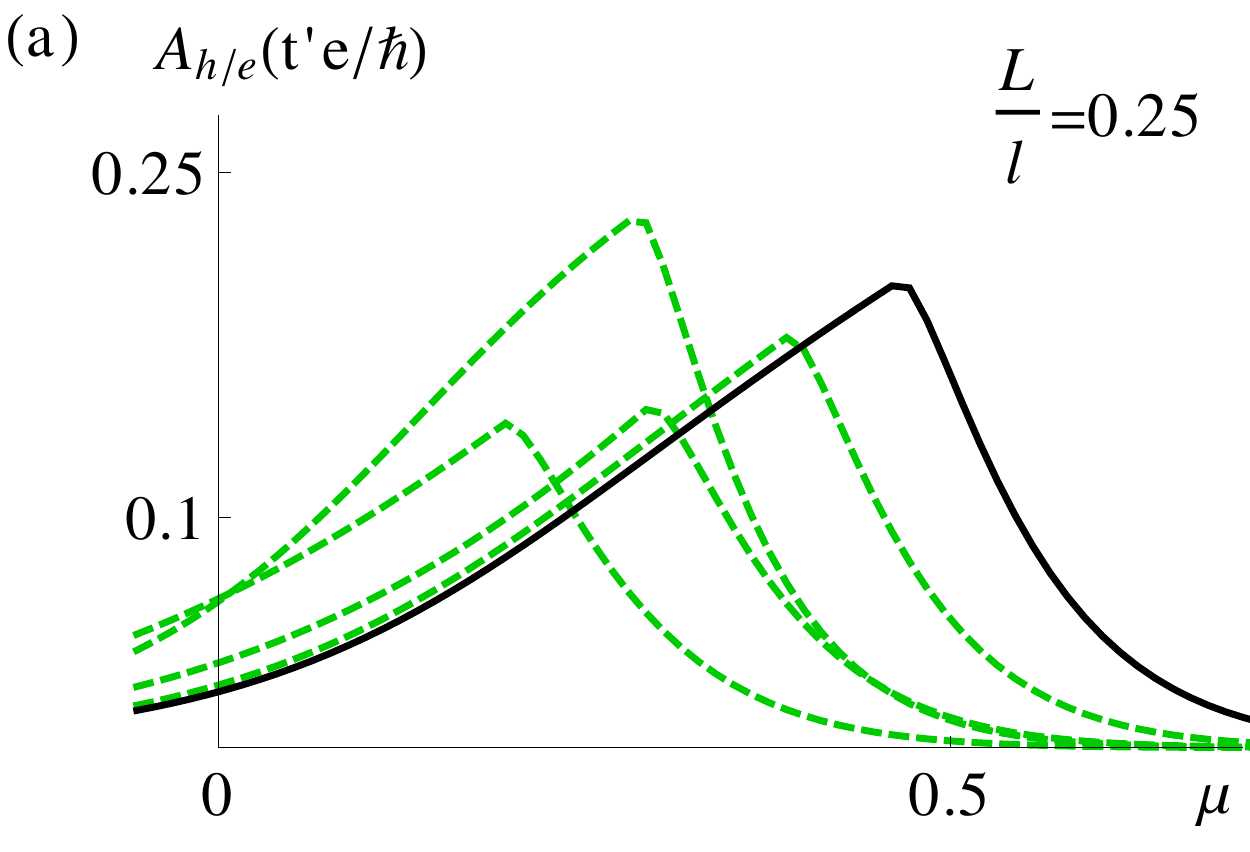}\\
\includegraphics[width=.46\textwidth]{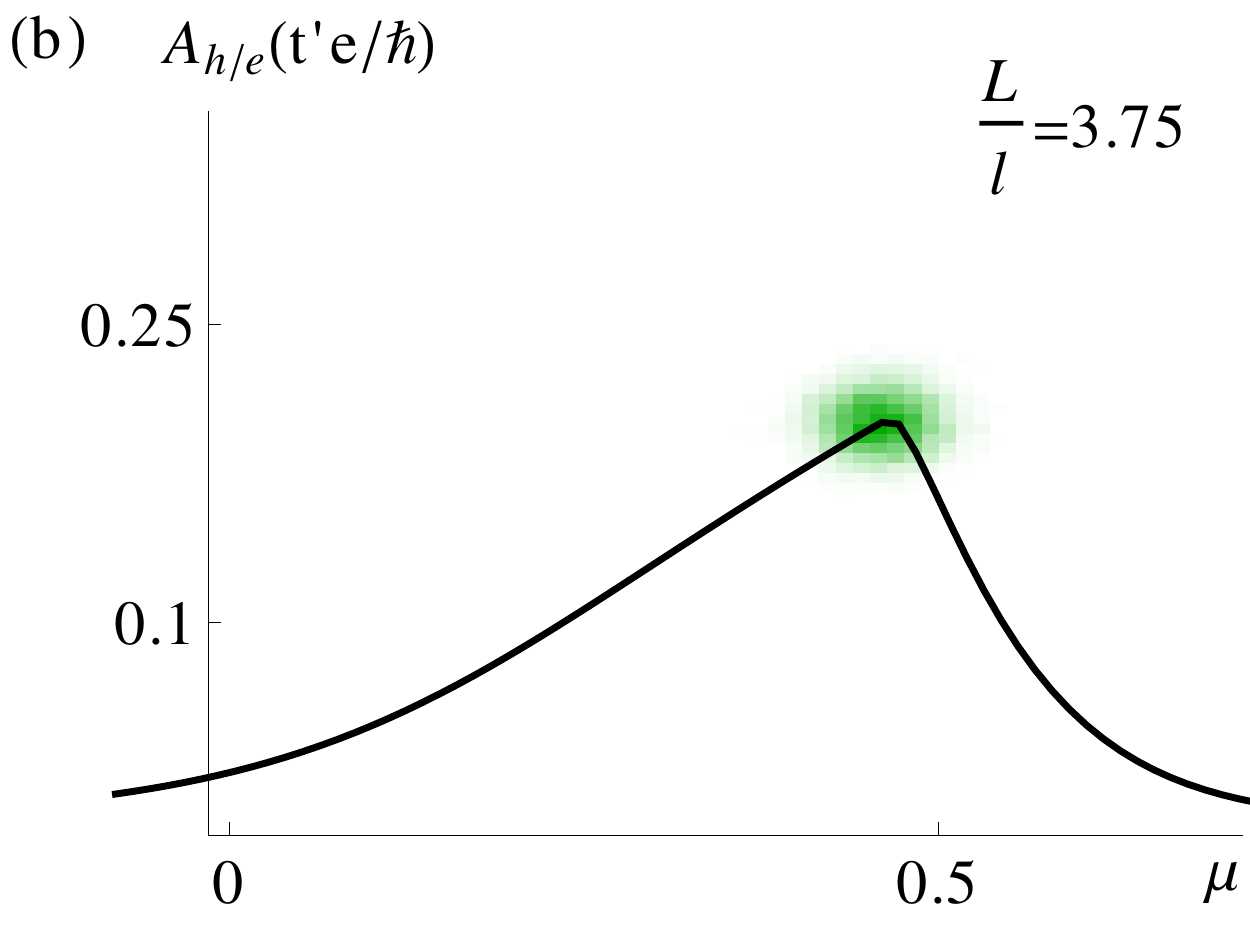}
\includegraphics[width=.46\textwidth]{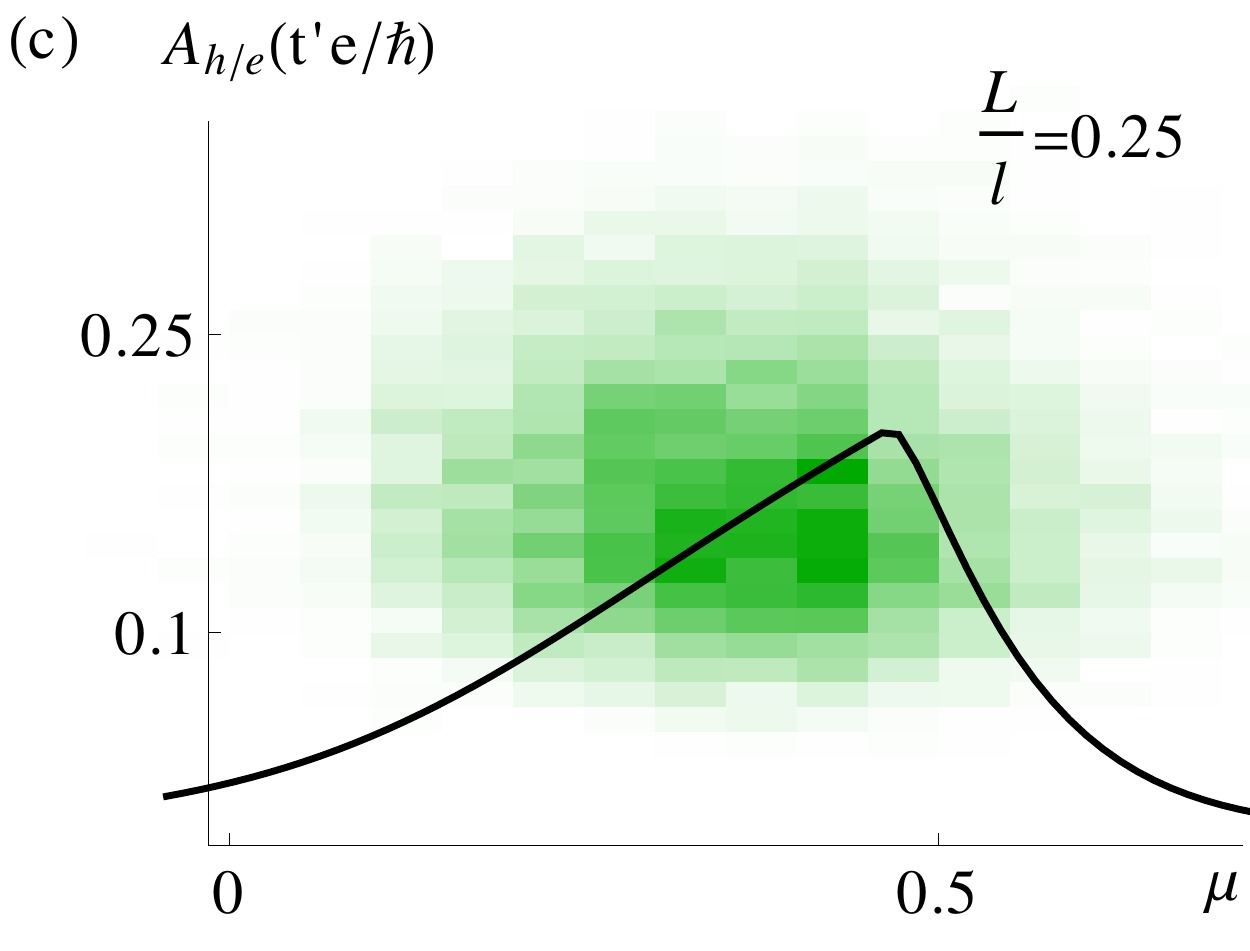}
\end{center}
\caption{(Color online) (a) Numerical results for the $h/e$-periodic Fourier component of the Josephson current, $A_{h/e}$, as function of chemical potential for a clean ring (black solid line) and disordered rings with four disorder configurations (green dashed lines) corresponding to $l=5$. (b) $A_{h/e}$ for a clean wire (black solid line) together with the histogram of the peak position in the presence of disorder for $l=75$ as a color code (green (gray) area).
(c) Same as (b) with $l=5$. For all plots we chose $L=20$ and $\Delta_{\rm TB}=1$.}
\label{fig:histres}
\end{figure}

In this section we investigate the fate of the peak in the equilibrium $h/e$-periodic Josephson current in the presence of disorder. Our main results are:
\begin{enumerate} 
\item[(i)] The typical peak height is not affected by disorder as long as the mean free path is longer than the circumference of the ring. Thus the signature persists in the presence of moderate disorder. 
\item[(ii)] For stronger disorder the peak height decreases and the peak position is shifted to lower chemical potentials.
\end{enumerate}

To study the effect of disorder we add a random onsite potential $\sum_i V_i c_i^\dagger c_i$ to the tight-binding Hamiltonian (\ref{Kitaev_TB_Hamiltonian}) and (\ref{Kitaev_TB_tunneling_Hamiltonian}), where the $V_i$ are taken from a uniform distribution over the interval $[-W,+W]$. The mean free path is then related to the disorder strength as $l\propto 1/W^2$ \footnote{For the numerical results for the tight-binding model we extract the mean free path from the variance of the normal distribution of $\ln(\epsilon_0)$ according to Eq.~(\ref{probab_distr_eps0}).}. To obtain numerical results we compute the spectrum by exact diagonalization. 
Disorder affects the $h/e$-periodic Josephson current by introducing fluctuations in the quantities $\epsilon_0$ and $\Gamma$. While $\Gamma$ is mainly affected by local fluctuations of the probability density of the Majorana wavefunction at the junction, $\epsilon_0$ fluctuates due to the disorder potential in the entire ring.
The interior overlap in disordered wires has been investigated previously for the continuum model (\ref{Kitaev_hamiltonian}) in regime (i), i.e., $\mu\gg m\Delta'^2$ \cite{Brouwer2011a}, where disorder leads to an increase of $\epsilon_0$ and subsequently to a disorder-induced phase transition to the nontopological phase.

Fig.~\ref{fig:histres}a shows numerical results for the $h/e$-periodic Josephson current for a few disorder configurations. The peaks in the presence of disorder (green dashed curves) are of comparable height as the peak in the clean ring (black solid curve). The peak shifts as a function of chemical potential which indicates fluctuations of the coherence length due to disorder. 

Surprisingly, the peak shifts to lower chemical potentials, corresponding to a decrease in $\epsilon_0$ with disorder in stark contrast to the known case of large $\mu$. This implies that the topological phase is {\em stabilized} by disorder if the system is close to the phase transition.
To investigate this further we plot the height and position of the peak maxima of many disorder configurations as a color code histogram for $l>L$ in Fig.~\ref{fig:histres}b and $l<L$ in Fig.~\ref{fig:histres}c. Indeed the average peak height is comparable to the one in the clean case for $l>L$. When $l\lesssim L$ the average peak height starts to decrease.
The histogram in Fig.~\ref{fig:histres}c confirms that the peak is shifted to lower chemical potentials on average.

\begin{figure}[t!]
 \begin{center}
\includegraphics[width=.23\textwidth]{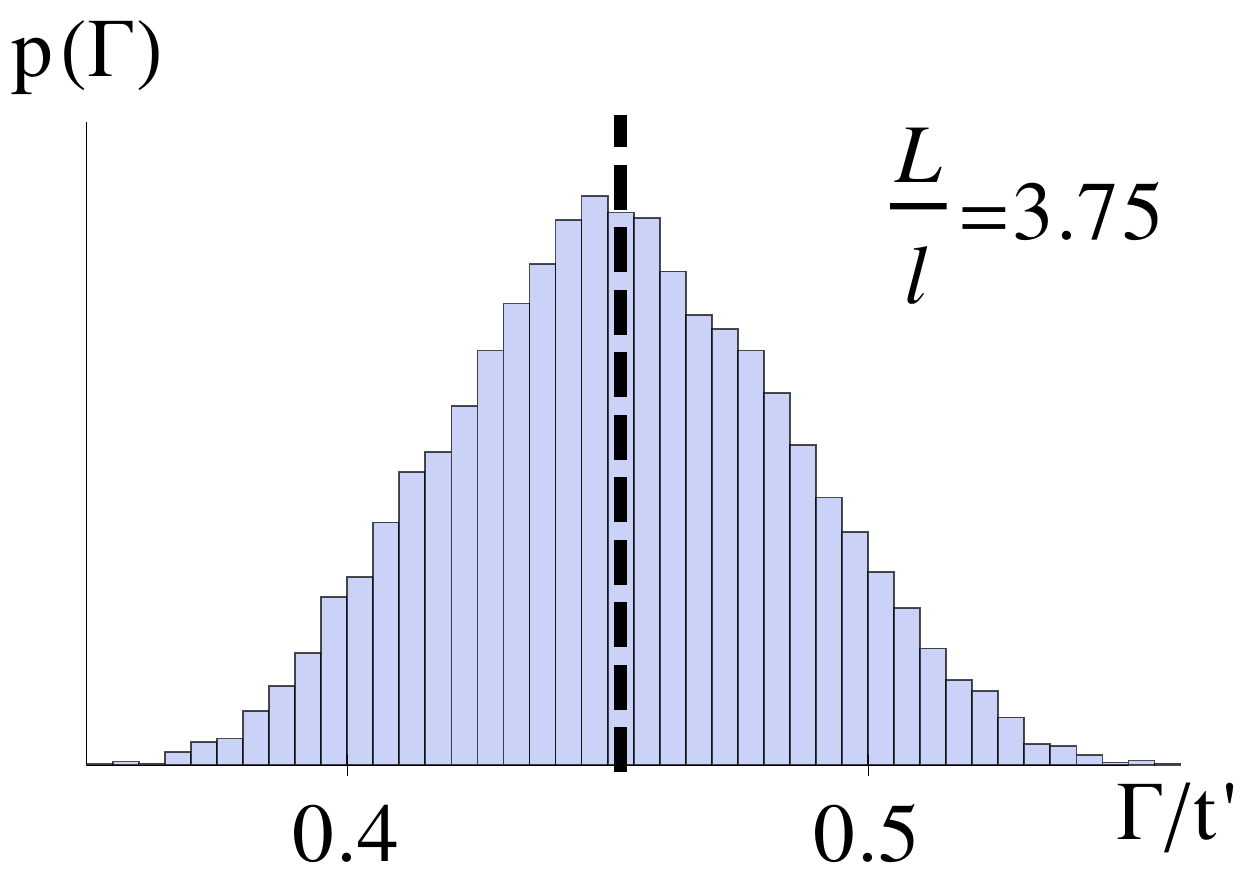}
\includegraphics[width=.23\textwidth]{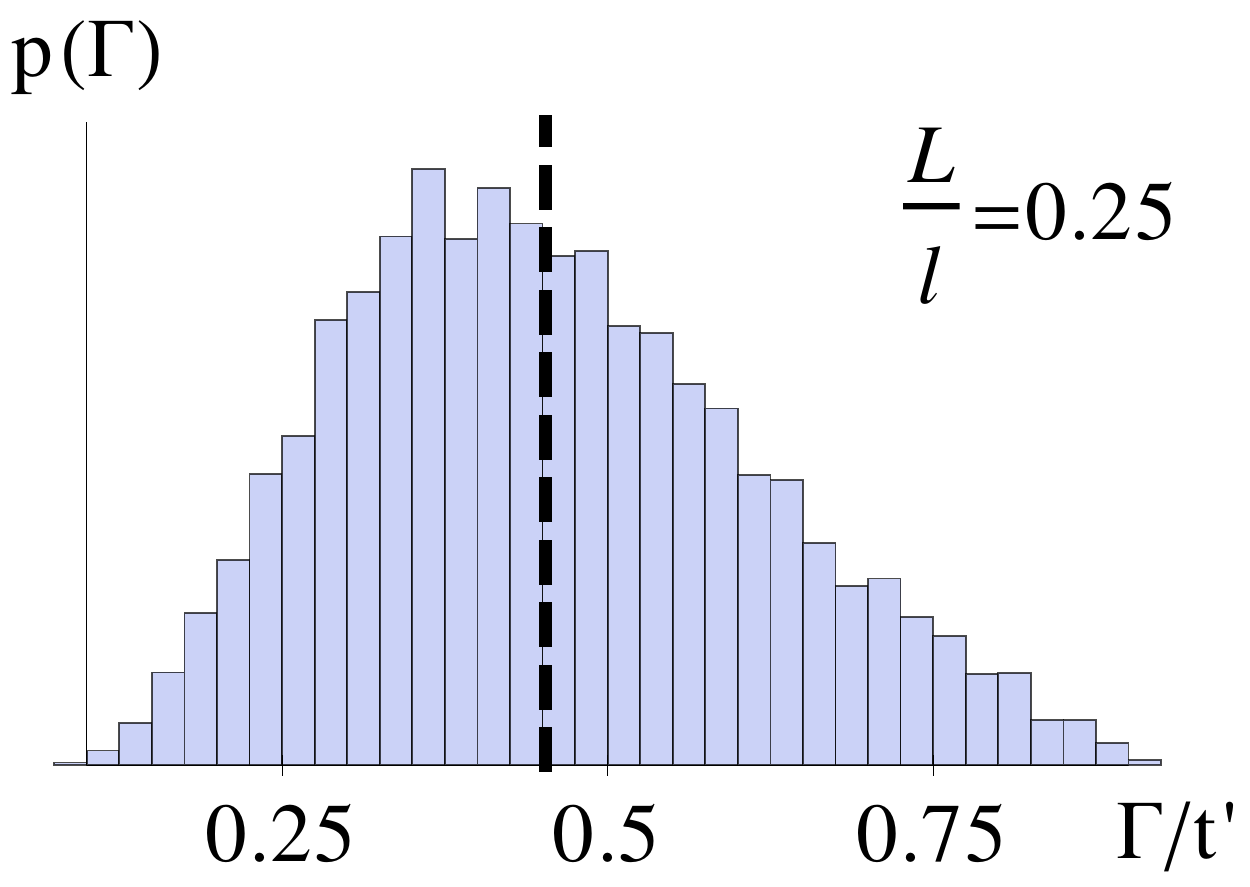}
\end{center}
\caption{(Color online) Histogram of $\Gamma$ for the same parameters as in Figs.~\ref{fig:histres}b and c at $\mu=0.4$. The dashed line denotes the value of $\Gamma$ for the clean ring.}
\label{fig:histgam}
\end{figure}

To understand this behavior we analyze the probability distributions of $\epsilon_0$ and $\Gamma$ over the disorder ensemble. In Fig.~\ref{fig:histgam} we show numerical results for the histogram of $\Gamma$ corresponding to the two ensembles in Figs.~\ref{fig:histres}b and c at $\mu=0.4$. For weak disorder the distribution is symmetric with a mean near the zero-disorder tunnel coupling. For larger disorder when $l<L$ the distribution becomes wider and asymmetric and the average decreases.

In order to determine the probability distribution for $\epsilon_0$ we now turn to the continuum Hamiltonian (\ref{Kitaev_hamiltonian}) for a wire of length $L$ without tunnel junction. To model short-range correlated disorder in the continuum model, we include a disorder potential with zero average $\langle V(x)\rangle =0$ and correlation function $\langle V(x) V(x^\prime)\rangle = \gamma \delta(x-x^\prime)$. For this model we employ a numerical method based on a scattering matrix approach \cite{Brouwer2003,Bardarson2007,Brouwer2011}. From the scattering matrix $S$ we obtain the lowest energy eigenstate $\epsilon_0$ by finding the roots of $\mathrm{det}(1-S(\epsilon))$.
In this model, the probability distribution of the hybridization energy $\epsilon_0$ has been shown to be log-normal in Ref.~\onlinecite{Brouwer2011a}. Specifically, it was shown that the log-normal distribution is governed by 
\begin{align}
\begin{split}
\Braket{\ln\left(\epsilon_0/2\Delta_{\rm eff}\right)}&=-L\left(\frac{1}{\xi}-\frac{1}{2l}\right)\\
 \text{var} \ln\left(\epsilon_0/2\Delta_{\rm eff}\right)&=\frac{L}{2l}.
 \end{split}
\label{langevin_kitaev}
\end{align}
for regime (i). The distribution function reflects the disorder-induced phase transition to the nontopological state at $\xi = 2l$.

\begin{figure}[t!]
 \begin{center}
\includegraphics[width=.48\textwidth]{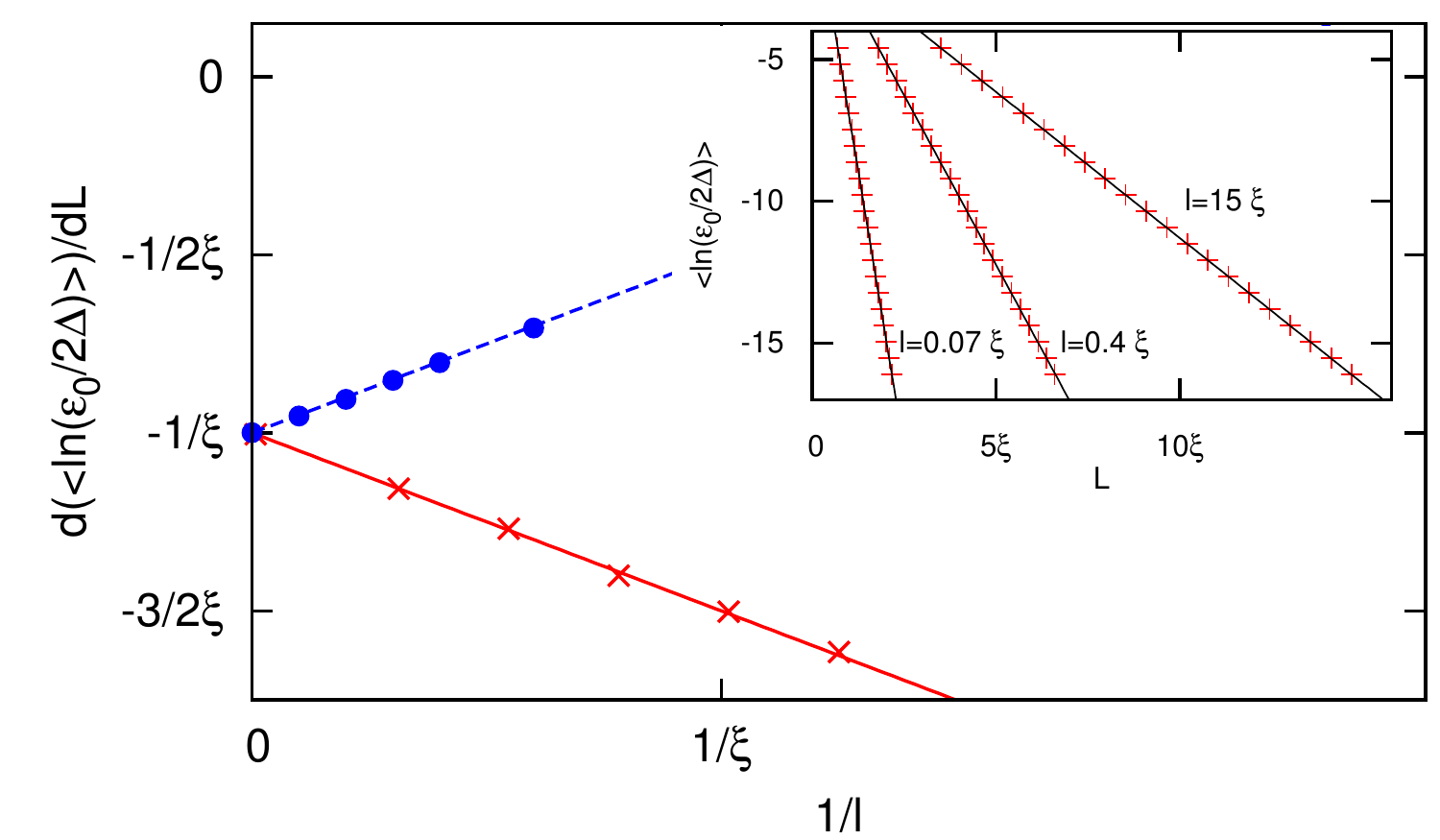}
\end{center}
\caption{(Color online) Slope of $\Braket{\ln(\epsilon_0(L)/2\Delta_{\rm eff})}$ (see inset) vs. disorder strength $1/l$ for regimes (i) with $\mu=300$ m$\Delta'^2$ (blue dots) and (ii) with $\mu=3\times 10^{-3} $ m$\Delta'^2$ (red crosses) together with theoretical prediction according to Eqs.~(\ref{langevin_kitaev})  (dashed line) and (\ref{langevin_dirac}) with $\lambda=1/2$ (solid line). Inset: numerical data (red crosses) and linear fit (solid) of the average of $\ln\epsilon_0$ as a function of $L$ for different disorder strengths.}
\label{fig:slope}
\end{figure}

The numerical results are presented in Fig.~\ref{fig:slope}. In the inset, we show that the mean of $\ln(\epsilon_0/2\Delta_{\rm eff})$ is indeed linear in $L$ with the slope depending on disorder strength. This slope is plotted as a function of inverse mean free path in Fig.~\ref{fig:slope}. The data for $\mu=300 m\Delta'^2$ (blue dots) agrees well with the prediction Eq.~(\ref{langevin_kitaev}) with the definitions $l=v_F^2/\gamma$ and $\xi=1/m\Delta'$.

The same plot also shows data corresponding to regime (ii), i.e., $\mu\ll m\Delta'^2$, marked by red crosses. Here, we have $l=\Delta'^2/\gamma$ and $\xi=\Delta'/\mu$. Clearly, the behavior is qualitatively different from regime (i), since disorder decreases $\epsilon_0$ rather than increasing it. This is consistent with the shift of the peak of $A_{h/e}$ to lower $\mu$.

In order to gain analytical insight we now derive the probability distribution of $\epsilon_0$ in regime (ii) extending the results of Ref.~\onlinecite{Brouwer2011a}. The relevant momenta at low energies in this regime are near $p=0$ (cf.\ Fig.~\ref{fig:spectrum}b). Linearizing the dispersion around this point yields the Dirac Hamiltonian Eq.~(\ref{Dirac_Hamiltonian}), where the disorder potential enters as a random mass term. 
Since the disorder potential is short-range correlated it couples high- and low-momentum degrees of freedom in the original Hamiltonian. Thus a proper linearization of the Hamiltonian requires one to project out the high-momentum states, which renormalizes the gap.

For a strictly linear model with a random mass term, the overlap $\epsilon_0$ has a log-normal distribution \cite{Brouwer2011a},
\begin{align}
\begin{split}
 \Braket{\ln\left(\epsilon_0/2\Delta_{\rm eff}\right)}&=-\frac{L}{\xi},\\
 \text{var} \ln\left(\epsilon_0/2\Delta_{\rm eff}\right)&=\frac{L}{l}.\end{split}
\label{langevin_dirac}
\end{align}
Thus for the Dirac Hamiltonian the mean of $\ln\left(\epsilon_0/2\Delta_{\rm eff}\right)$ does not depend on disorder. 
A systematic linearization of the disordered spinless $p$-wave superconductor in the vicinity of the topological phase transition effectively renormalizes the chemical potential $\mu$ and hence the coherence length $\xi=\Delta'/\mu$. 

We start by defining the projection operators $P=\sum_{|p|<p_1}\Ket{\psi_p}\Bra{\psi_p}$ onto the low momentum subspace and $Q=1-P$, where $\lbrace\Ket{\psi_p}\rbrace_p$ is a complete set of momentum eigenstates of the clean Hamiltonian. 
The relevant momentum scale for this projection is given by $p_1=m\Delta'$, since for $p\ll p_1$, the term $p^2/2m$ constitutes the lowest energy scale of the Kitaev Hamiltonian.
Furthermore, we assume that the disorder potential does not affect high momenta $p_1\gg 1/l$.
We can now project the clean Kitaev Hamiltonian $H$ to the low- and high-energy subspaces,
\begin{align}
PHP&\simeq P\left[(-\mu+V(x))\tau_z+\Delta'p\tau_x\right]P,\\
QHQ&\simeq Q\left(p^2/2m\right)\tau_zQ.
\end{align}
Both subspaces are exclusively mixed by the disorder potential $PHQ=PV(x)\tau_zQ$. To second order in $V$, the correction to the low-energy Hamiltonian is then given by
\begin{align}
 \delta H(p)&\simeq \Braket{\psi_p|PHQ\left(\epsilon_p-QHQ\right)^{-1}QHP|\psi_p}\nonumber\\
&\simeq\sum_{|p'|>p_1}V_{pp'}\frac{1}{\epsilon_p-p'^2/2m\tau_z}V_{p'p}.
\end{align}
Here we used the short notation $V_{pp'}=\Braket{\psi_p|V(x)|\psi_{p'}}$. Averaging over disorder, we obtain
\begin{align}
\Braket{ \delta H(p)}&\simeq-\sum_{|p'|>p_1}\frac{2m\gamma}{p'^2}\tau_z\sim-\frac{\gamma}{\Delta'}\tau_z.
\end{align}

Thus the renormalization produces a contribution to the low-energy Hamiltonian which has the same structure as the chemical potential term. Hence we find a renormalized chemical potential
$\mu'=\mu+\lambda\gamma/\Delta'$ with a numerical factor $\lambda>0$ that cannot be determined from this argument. Thus disorder enters the final result through the renormalized coherence length
\begin{align}
 \frac{1}{\xi}\rightarrow \frac{1}{\xi}+\frac{\lambda}{l}.\label{renormalized_coherence_length}
\end{align}
The data in Fig.~\ref{fig:slope} confirm Eqs.~(\ref{langevin_dirac}) and (\ref{renormalized_coherence_length}) and determine the unknown numerical prefactor to be $\lambda=1/2$. 
Thus for $\mu\ll m\Delta'^2$, $\epsilon_0$ has a log-normal distribution with mean and variance given by
\begin{align}
\begin{split}
 \Braket{\ln\left(\epsilon_0/2\Delta_{\rm eff} \right)}&=-L\left(\frac{1}{\xi}+\frac{1}{2l}\right),\\
 \text{var} \ln\left(\epsilon_0/2\Delta_{\rm eff}\right)&=\frac{L}{l}.\end{split}\label{probab_distr_eps0}
\end{align}
This result is very similar to Eq.~(\ref{langevin_kitaev}) where, however, the disorder correction to the decay length enters with opposite sign. This underlines the contrast between the two regimes, i.e., that 
disorder drives the system further into the topological phase when it is close to the phase transition, but away from it for larger chemical potentials.
Specifically a spinless $p$-wave superconducting wire with negative chemical potential may exhibit edge states with an energy exponentially small in $L$ as long as disorder is strong enough.

Combining the disorder-induced fluctuations of $\Gamma$ and $\epsilon_0$ we can understand the suppression of the peak in the $h/e$-periodic Josephson current in Fig.~\ref{fig:histres}c for $l< L$. While $\epsilon_0$ is decreased on average for a given $\mu$ with increasing disorder, $\Gamma$ does not increase at the same time and thus the average peak height decreases. 
However the fluctuations of $\Gamma$ and $\epsilon_0$ become larger as disorder increases such that for single disorder configurations significant peaks are still possible even if the average peak height decreases.

\subsection{Phase diagram of a disordered wire}
\begin{figure}[t!]
\includegraphics[width=.48\textwidth]{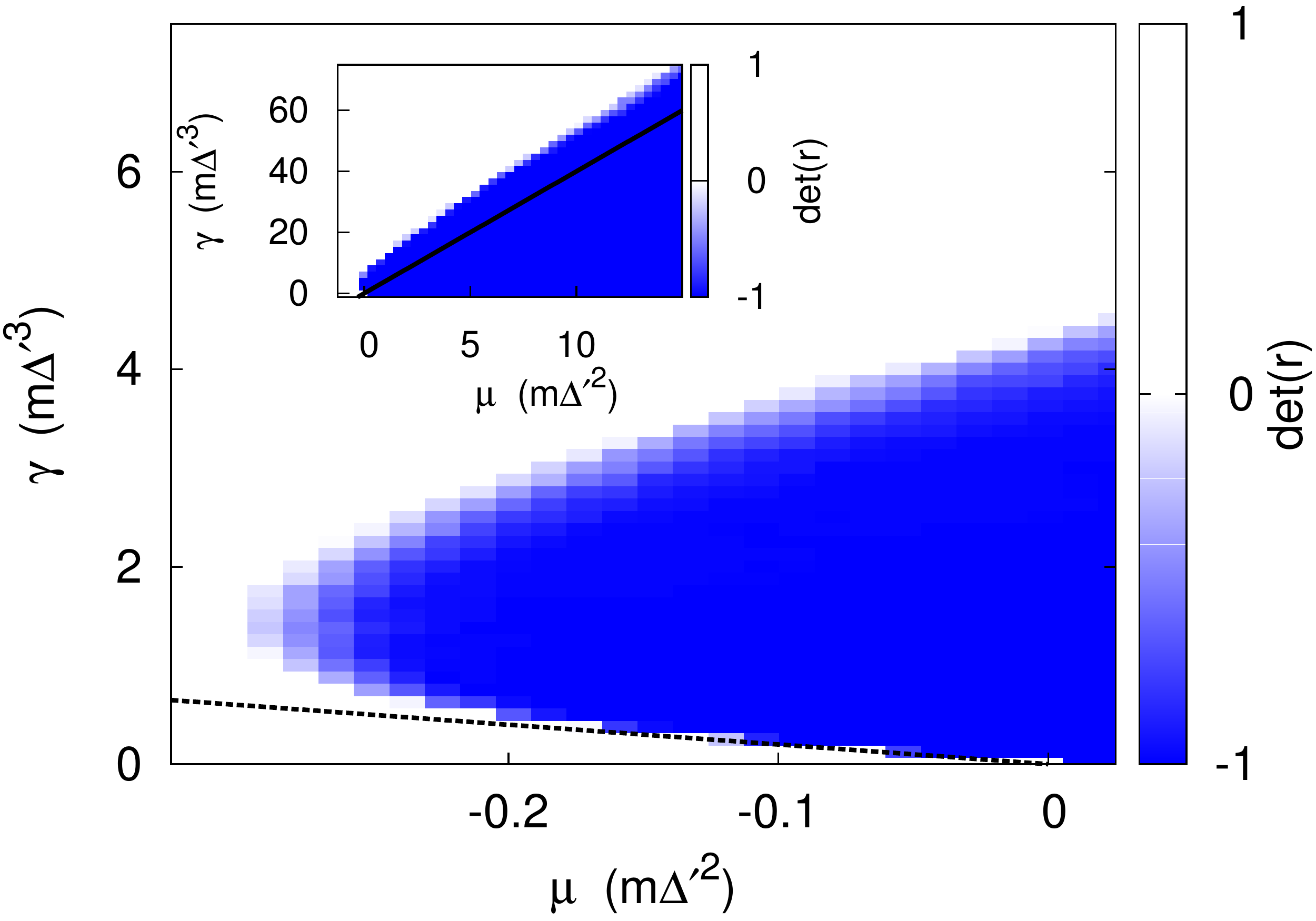}
\caption{(Color online) Phase diagram of the continuum model (\ref{Kitaev_hamiltonian}) as function of disorder strength $\gamma$ and chemical potential $\mu$ in the regime $\mu\ll m\Delta'^2$. The data has been averaged over 100 disorder configurations. For $\mu<0$ disorder gives rise to a trivial-to-topological phase transition with a reentrant nontopological phase for stronger disorder. The dashed line denotes the phase transition line $\gamma^{(ii)}_c(\mu)$ valid for small $|\mu|$ given in Eq.~(\ref{phase_boundary}). Inset: Phase diagram for a larger range of $\mu$ and $\gamma$. The solid line represents the predicted phase boundary $\gamma^{(i)}_c(\mu)$ for large $\mu$. (The analytical phase boundary is only accurate at large $\mu$ up to sublinear corrections.)}
\label{fig:phase_diagram}
\end{figure}

Motivated by the contrasting probability distributions of $\epsilon_0$ in the regimes of large and small $\mu$ we numerically calculate the phase diagram of the continuum model (\ref{Kitaev_hamiltonian}) as a function of $\mu$ and $\gamma$, particularly paying attention to the region near the topological phase transition of the clean model.
By means of the scattering matrix approach also used in the last section we compute the determinant of the reflection matrix of a wire of length $L$ at $\epsilon=0$ which approaches the values $+1$ and $-1$ as $L\rightarrow\infty$ in the nontopological and topological phase, respectively \cite{Merz2002,Akhmerov2011}.

The resulting phase diagram is plotted in Fig.~\ref{fig:phase_diagram}. From Eq.~(\ref{langevin_kitaev}) we infer that the topological phase transition occurs for $\xi=2l$ in the regime $\mu\gg m\Delta'^2$.
Using the definitions of $l$ and $\xi$ in this regime, we obtain the phase boundary $\gamma_c^{(i)}(\mu)=4\Delta'\mu$.
This is compared with the numerical results in the inset of Fig.~\ref{fig:phase_diagram}. 
The numerically calculated phase boundary $\gamma_c^{\rm num}(\mu)$ has only sublinear deviations from the predicted line, so that the ratio $\gamma_c^{\rm num}(\mu)/2\mu\Delta'=\xi/l$ approaches the value $2$ for $\mu\rightarrow\infty$ as expected.

However, near $\mu=0$ the behavior is qualitatively different. Here, disorder can induce a topological phase for $\mu<0$ as well as a reentrant nontopological phase at larger disorder. 
From Eq.~(\ref{probab_distr_eps0}) we find the condition $\xi=-2l$ for the phase boundary. This corresponds to
\begin{align}
\gamma^{(ii)}_c(\mu)=-2\Delta'\mu,\label{phase_boundary} 
\end{align}
which we find to agree well with the numerical results for the phase diagram (see dashed line in Fig.~\ref{fig:phase_diagram}). Thus the phase diagram confirms that weak disorder leads to an enhancement of the chemical potential range of the topological phase, while for stronger disorder the range decreases again.

\section{Conclusion}

Even for conventional superconducting phases, the flux periodic currents
have been widely studied for mesoscopic rings \cite{Buttiker1986,Oppen1992,Schwiete2010,Koshnick2007}.
Here, we studied the Josephson currents across a weak link in a mesoscopic
ring in a topological superconducting phase. As a paradigmatic model
system, we studied Kitaev's model of a one-dimensional spinless $p$-wave
superconductor, focusing on the parameter regime near the topological
phase transition. We found that in mesoscopic rings, there is an
$h/e$-periodic contribution to the tunneling current even if electron
number parity is not conserved. This $h/e$-periodic contribution emerges
due to the hybridization of the Majorana bound states localized on the two
sides of the weak link through the interior of the ring and exhibits a
pronounced peak just on the topological side of the topological phase
transition. This peak provides an interesting signature for the existence
of a topological phase transition and the formation of Majorana fermions
at the junction.

We found that this effect remains robust in the presence of disorder in
the wire. In fact, near the topological phase transition disorder can even
stabilize the topological phase. When tuning, say, the chemical potential
of the system to the nontopological side of the phase transition, there is
a disorder-induced topological phase for moderate amounts of disorder,
with a reentrant nontopological phase at even stronger disorder. This is
in stark contrast to the behavior of the system far in the topological
phase where disorder weakens and eventually destabilizes the topological
phase.

\begin{acknowledgments}
We would like to acknowledge discussions with P.\ Brouwer, A. Haim, N.\ Lezmy, and G.\ Refael. We are grateful for partial support by SPP 1285 of the Deutsche Forschungsgmeinschaft (FvO and YO), the Virtual Institute ``New states of matter and their excitations'' (FvO), grants of ISF and TAMU (YO) as well as a scholarship of the Studienstiftung d.\ dt.\ Volkes (FP).
\end{acknowledgments}

\clearpage

\onecolumngrid
\begin{appendix}

\section{Effective energy splitting and Josephson coupling of Majorana
end-states}
\label{appendice}

We present here the analytical estimation of the
Majorana energy splitting, $\epsilon_0$, and the effective Josephson 
coupling, $\Gamma$, in the Hamiltonian (\ref{low_energy_Hamiltonian}). 
For sake of simplicity we will first present our calculation for the
effective Josephson coupling $\Gamma$, obtained by working in the
tight-binding model. We will then describe the calculation of
$\epsilon_0$ through an alternative approach working directly in the
continuum limit both in both regimes $\mu \ll m \Delta'^2$ and $\mu
\gg m \Delta'^2$ discussed in the main text.

\begin{figure}[hbt]
\includegraphics[width=.6\textwidth]{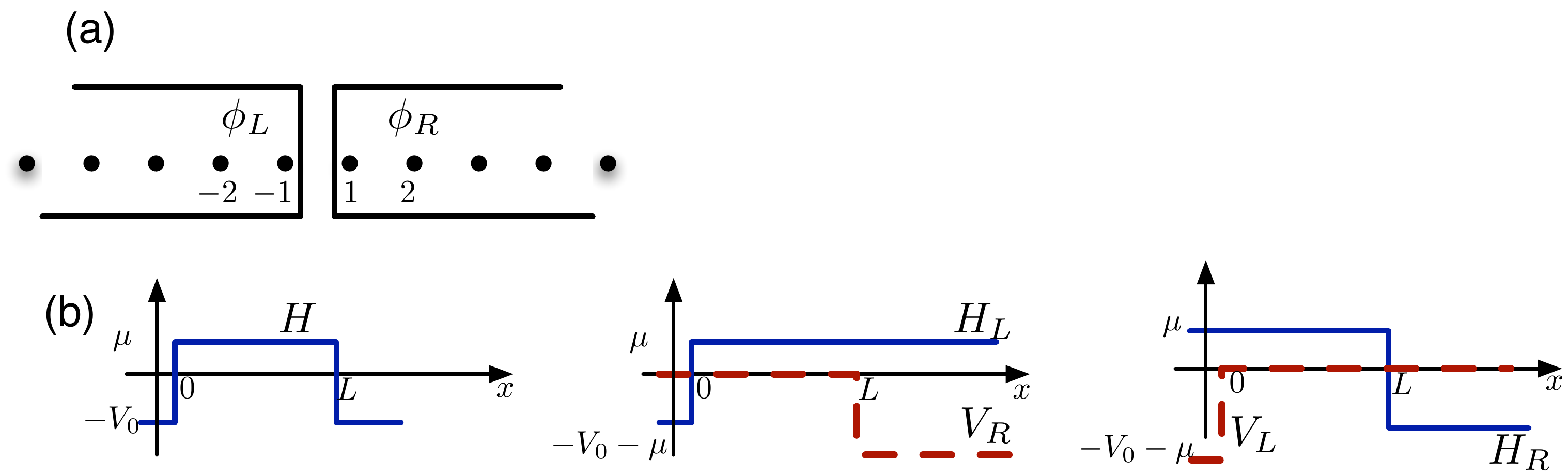}
\caption{(a) Sketch of the Josephson junction between two topological
  segments of the wire and spatial dependence of the gate potential in
  the corresponding continuum realization. (b) Spatial dependence of
  the chemical potential, $\mu(x)$, in the Hamiltonian of the finite length wire and
  for the approximate Hamiltonians, $H_L$ and $H_R$, used in the perturbative calculations}\label{ring-approx}
\end{figure}

\textbf{Effective Josephson coupling.}
In order to compute the effective Josephson coupling we neglect the
effect of the overlap of the two Majorana wavefunctions in the
topological part of the ring. 
We therefore consider the limit of a junction between two
half-infinite topological sectors of the wire, the right one on sites
$j \in [1,\infty)$ and the left one on $j \in (\infty,-1)$.
They are both described by the Hamiltonian (\ref{Kitaev_TB_Hamiltonian}) with paring strength
$\Delta_{\rm TB} e^{i\phi_{a}}$ with $a=L,R$ for the left
and right sector, respectively.
The hopping between the two sectors from Eq.~(\ref{Kitaev_TB_tunneling_Hamiltonian}) now simply reads
$H_{\textrm{T}}= -t' (c_{-1}^{\dag} c_1 +c_{1}^{\dag} c_{-1})$---cf.\ Fig.~\ref{ring-approx}. 

We include the effect of the tunneling between the two wires
perturbatively. 
The low energy excitations of this model for $t'=0$ are represented by
left and right zero energy Majorana operators \cite{Kitaev2001}
\begin{align}
b_{R} =A \sum_{j=1}^\infty (x_+^j-x_-^j)\gamma_{B,j}^{(R)} \\
b_{L}= A \sum_{j=-1}^{-\infty} (x_+^{-j}-x_-^{-j})\gamma_{A,j}^{(L)}
\end{align}
where $x_{\pm} = (- \mu_{\textrm{TB}} \pm \sqrt{\mu_{\textrm{TB}}^2
  -4t^2 + 4 \Delta_{\rm TB}^2 }) / (2t +2 \Delta_{\rm TB})$, $A=(\sum_{j} |x_+^j-x_-^j|^2)^{-1/2}$ is a normalization
constant, and the operators $\gamma_{A(B),j}$ are defined via
 $c_J=(\gamma_{B,j}^{(R)}e^{-i\phi_R/2}+i\gamma_{A,j}^{(R)}e^{i\phi_R/2})/2$.
The projection of $H_{\textrm{T}}$ onto the subspace spanned by the operators $b_R$, $b_L$ leads to the effective coupling between the Majorana states. The
Hamiltonian can be rewritten in terms of ordinary fermion operators $d_M=(b_R+ib_L)/2$
to take the form presented in Eq.~(\ref{effective_tunneling_hamiltonian}) with 
$\Gamma= t' A^2 |x_+ -x_-|^2$. For simplicity, we consider $\mu<2t\left(1-\sqrt{1-\Delta_{\rm TB}^2/t^2}\right)$ which corresponds to the condition $\mu<m\Delta'^2$ in the continuum model (\ref{Kitaev_hamiltonian}) and ensures that $x_{\pm}$ are real (see discussion of the continuum model in Sec.~\ref{sec:kitaev_basics}).
Explicitly one obtains
\begin{align}
\Gamma=\frac{ t'\mu (4t-\mu) \Delta_{\mathrm{TB}} }{ t(t+\Delta_{\mathrm{TB}})^2} 
\label{good-for-plots}
\end{align}
In the continuum limit, when $\mu \ll\Delta_{\mathrm{TB}} \ll t$ the expressions simplifies to
\begin{align}
\Gamma \approx 4 t' \mu \Delta_{\mathrm{TB}} / t^2 \, .
\end{align}

\textbf{Majorana energy splitting.}
In order to compute the energy splitting $\epsilon_0$ we employ an
alternative method working directly in the continuum limit.
Similar to before, we neglect here the  Majoranas interaction through the Josephson
junction.
The problem is then completely equivalent to calculating the energy
splitting of two Majorana at the end of a wire of length $L$.
We start considering the regime (ii) described by the Hamiltonian in  Eq.~(\ref{Dirac_Hamiltonian}), $H =
\mu(x) \tau_z +\Delta'(x) p \tau_x$,
with the specific choice of parameters (cf.\ Fig.~\ref{ring-approx})
\begin{eqnarray}
\label{esatta}
& & \mu(x) = -V_0 [\T{-x}+\T{x-L}] +\mu \T{x}\T{L-x} \, , \nonumber\\
& & \Delta'(x) =\Delta' \, ,
\end{eqnarray}
where $V_0 >0$ and $\mu >0$ guarantee that the wire is in a topological
phase in $[0,L]$ and in a nontopological phase otherwise. 

We determine the Majorana wavefunctions by a perturbative approach. 
We first find the wavefunctions
of the Majorana fermion localized at one of the interfaces between the 
topological and the insulating region, thus fully ignoring the
existence of the other Majorana state. 
We label the corresponding states at the interfaces at $x=0$ and $x=L$ as $\ket{L}$ and
$\ket {R}$ respectively. 
We then project the Hamiltonian onto the Majorana subspace, to
obtain their effective interaction,
$H_{\textrm{overlap}}= \sum_{\alpha,\beta=L,R} \langle \alpha |H
|\beta\rangle \, \ket{\alpha} \bra{\beta}$.
Eventually we will be interested in the limit $V_0 \rightarrow \infty$
corresponding to a high insulating barrier outside the wire.

The Majorana state at $x=0$ is the zero-energy eigenstate of the
Hamiltonian $H_L$, defined again by the same Hamiltonian as in Eq.~(\ref{Dirac_Hamiltonian}), but
with the choice of parameters
\begin{eqnarray}
\label{sinistra}
& & \mu(x) = -V_0 \T{-x} +\mu \T{x} \, , \nonumber\\
& & \Delta'(x) =\Delta' \, ,
\end{eqnarray}
as depicted in Fig.~\ref{ring-approx}.
Solving this equation separately for $x>0$ and $x<0$ leads to zero
energy states with imaginary momenta.
Namely we can write
\begin{equation}
\label{majol}
\ket{L} = \mathbf{v}_{-,L} e^{\alpha_0 x} \T{-x} + \mathbf{v}_{+,L} e^{-\alpha
  x} \T{x} \, 
\end{equation} 
where $\alpha= \mu /|\Delta'|$ is the inverse coherence length in the
wire, and $\alpha_0= V_0/|\Delta'|$.
The  twodimensional vectors $\mathbf{v}_{\pm, L}$ are determined by the
continuity of the wavefunction and its derivative at the interface
and by the wavefunction normalization. They are given by:
\begin{equation}
\mathbf{v}_{+,L}=\mathbf{v}_{-,L}= \sqrt{\frac{\alpha \alpha_0}{\alpha+
    \alpha_0}} \left( \begin{array}{c} 1 \\ -i \end{array} \right) \, .
\end{equation}
In the limit $V_0 \rightarrow \infty$ we are interested in, they
reduce to
\begin{equation}
\mathbf{v}_{+,L}=\mathbf{v}_{-,L} = \sqrt{\alpha} \left( \begin{array}{c} 1 \\ -i \end{array} \right) \, .
\end{equation}
In a completely analogous way we can calculate the zero-energy
eigenstate of the Hamiltonian $H_R$ defined once more by the Hamiltonian
in Eq.~(\ref{Dirac_Hamiltonian}), now with (cf.\ Fig.~\ref{ring-approx})
\begin{eqnarray}
& & \mu(x) = -V_0 \T{x-L} +\mu \T{L-x} \, ,\\
& & \Delta'(x) =\Delta' \, .
\end{eqnarray}
In this case the zero-energy eigenstate reads
\begin{equation}
\label{major}
\ket{R} = \sqrt{\frac{\alpha \alpha_0}{\alpha+
    \alpha_0}} \left( \begin{array}{c} 1 \\ i \end{array} \right) 
\left[ e^{-\alpha_0 (x-L)} \T{x-L} + e^{\alpha (x-L)} \T{L-x}  \right] \, . 
\end{equation}
In the limit $V_0 \rightarrow \infty$ the prefactor reduces to
$\sqrt{\alpha}$. Note that the particle-hole superposition has
different phases in $\ket{L}$ and $\ket{R}$. 

We can then project the full Hamiltonian onto the low-energy subspace
spanned by the two Majorana states. 
In doing so, we conveniently
rewrite $H=H_L +V_R =H_R +V_L$, 
where $V_R = - (V_0+\mu) \T{x-L} \tau_z$ and $V_L =- (V_0+\mu) \T{-x}
\tau_z$, and the spatial dependence of the various terms is presented in Fig.~\ref{ring-approx}
We can then compute, e.g., $\langle L |H |  L \rangle = \langle L |H_L
|  L \rangle + \langle L |V_R |  L \rangle =\langle L |V_R |  L
\rangle$, and all the other matrix elements in a similar way, to get
\begin{equation}
H_{\textrm{overlap}} = \left( \begin{matrix} 
\langle L |V_R |  L \rangle & \langle L |V_L |  R \rangle \\
\langle R |V_L |  L \rangle & \langle R |V_L |  R \rangle 
\end{matrix}\right) = \left( \begin{matrix} 
0 & -2 \mu e^{- \alpha L} \\
-2 \mu e^{- \alpha L} & 0
\end{matrix}\right) \, .
\end{equation}
This leads to the energy splitting $\epsilon_0 =2 \mu e^{-\alpha L}$,
as reported in Section \ref{sec:low_energy_model}.

In complete analogy, one can perform the calculation of the energy
splitting in the regime $\mu \gg m \Delta'^2$.
In this case the appropriate Hamiltonian, $H$, is that in Eq.~(\ref{Kitaev_hamiltonian}),
with the same choice of parameters as in Eq.~(\ref{esatta}). 
Again we start considering the left interface, looking for zero-energy
solutions of the $H_L$, i.e., the Hamiltonian in Eq.~(\ref{Kitaev_hamiltonian}) with the choice of
parameters as in Eq.~(\ref{sinistra}). 
We introduce $a_0 = V_0/(m
\Delta'^2) >0$ and $a = \mu /(m \Delta'^2) > 0$. 
We can write the solutions as
\begin{eqnarray}
 &  &\ket{L} =\sqrt{m \Delta'}\sqrt{\frac{a}{1+2 a}}
\left( \begin{matrix} 1 \\ -i \end{matrix}\right) \left[ (1+\eta)
  e^{i\kappa_-x} \T{-x} + \left( e^{i k_+x} +\eta e^{i k_-x}\right) \T{x} \right]
\, , \\
 &  &\ket{R} =\sqrt{m \Delta'}\sqrt{\frac{a}{1+2 a}}
\left( \begin{matrix} 1 \\ i \end{matrix}\right) \left[  \left( e^{-i k_+(x-L)} +\eta e^{-i k_-(x-L)}\right) \T{L-x}+ (1+\eta)
  e^{-i\kappa_-x} \T{x-L}  \right] \, ,
\end{eqnarray}
where $\kappa_- = -i m|\Delta'| (\sqrt{2 a_0})$, $k_{\pm}=m
|\Delta'| (\pm\sqrt{2 a} +i)$ and $\eta= (\kappa_--k_+)/(k_--\kappa_-)$.
In all the expressions we are neglecting $\mathcal{O}(\sqrt{a}/\sqrt{a_0},
1/\sqrt{a_0})$, consistent with the condidtion $a_0 \gg a \gg 1$, reflecting
the limit $V_0 \rightarrow\infty$ and the regime under consideration.
The matrix elements of the effective Hamiltonian are, in this case, 
$\langle L |H |  L \rangle =\langle R |H |  R \rangle =0$ and $\langle
L |H |  R \rangle =\langle R |H |  L \rangle \equiv 
-\epsilon_0$ , with
\begin{equation}
\epsilon_0=\Delta' \sqrt{2m \mu} \left[ \sin ( \sqrt{2 m \mu} L) -
  \sqrt{a/a_0} (\sqrt{2} -1) \cos ( \sqrt2 {m \mu})
  L +\mathcal{O} (\sqrt{a} /a_0^2) \right] \, .
\end{equation}

\end{appendix}

\end{document}